\documentstyle[12pt,epsf]{article}
\begin{document}
\begin{center}
{\large\bf Proposal and theoretical formalism for studying baryon
radiative decays from $J/\psi \to B^*\bar B + \bar{B^*}B \to \gamma
B\bar B$}

\vspace{1cm} Sayipjamal Dulat$^{a,c,}$\footnote{dulat98@yahoo.com},
Jia-Jun Wu$^{b}$ and
B. S. Zou$^{b,c}$\\

a) School of Physics Science and Technology, Xinjiang
University,Urumqi, 830046, China\\
b) Institute of High Energy Physics, CAS,  P.~O.~Box 918(4), Beijing
100049, China\\
c) Theoretical Physics Center for Science Facilities, CAS,
Beijing 100049, China\\

\date{\today}
\end{center}
\begin{abstract}

 With accumulation of high statistics data at BESIII, one may
 study many new interesting channels. Among them, $J/\psi \to B^*\bar B + \bar{B^*}B
\to \gamma B\bar B$ processes may provide valuable information of
the radiative decays of the excited baryons $(N^*,\Lambda^*,
\Sigma^*, \Xi^*)$, and may shed light on their internal
 quark-gluon structure. Our estimation for the branching ratios of the
nucleon excitations $N^*(1440)$, $N^*(1535)$ and $N^*(1520)$ from
the reaction $J/\psi \to N^*\bar p + \bar N^* p \to p\bar p\gamma$,
indicates that these processes can be studied at  BESIII with
$10^{10}$ $J/\psi$ events. Explicit theoretical formulae for the
partial wave analysis (PWA) of the $J/\psi\to B^*\bar B +\bar B^* B$
with $B^*\to B\gamma$ and $\bar B^*\to \bar B\gamma$ within
covariant L-S Scheme are provided.
\end{abstract}


\section{Introduction}
Baryons $B(N,\Lambda, \Sigma, \Xi, \cdots)$ and their excited states
$B^*(N^*,\Lambda^*, \Sigma^*, \Xi^*, \cdots)$ are complex systems of
confined quarks and gluons. Excited baryons  are sensitive to
details of quark confinement~\cite{Isgur1} which is poorly
understood within the fundamental theory of strong interactions -
Quantum Chromodynamics (QCD). Thus, understanding their structure
and determining their properties (masses, decay widths, branching
ratios, spins, parities, electromagnetic form factors, magnetic
moments, polarizabilities) will provide a better understanding of
how confinement works in baryons. Concerning the internal
quark-gluon structure of baryons there are various proposed
configurations: (a) the classical constituent three quark $(qqq)$
states; (b) $qqqg$ hybrid states~\cite{Barnes1}; (c) diquark-quark
states~\cite{diquark,Zhaoq}; (d) meson-baryon
states~\cite{Thomas}-\cite{Oset1}; (e) pentaquark with diquark
clusters~\cite{Riska}-\cite{zhusl1}, etc. A series of new
experiments on excited nucleon $N^*$ physics with electromagnetic
probes have been started at modern facilities such as
TJNAF~\cite{Clas}, ELSA~\cite{Elsa}, GRAAL~\cite{Graal},
SPRING8~\cite{Spring8} and BEPC~\cite{BES1,BES2}. In last few years
these facilities provided a considerable amount of precise data for
various excited nucleon production and decay channels  and opened a
great opportunity to make quantitative investigations of the baryon
structure. To extract properties of $N^*$ resonances partial wave
analysis (PWA) is necessary. In this paper, first we show that the
radiative decays of baryons can be studied at BESIII with expected
$10^{10}$ $J/\psi$ events. Then we provide PWA formulae within
covariant L-S Scheme~\cite{Zou3} for multi-step chain processes
$J/\psi \to B^*\bar B + \bar{B^*}B \to \gamma B\bar B$. Because
electromagnetic transition rates of excited baryons to their
respective ground states offer a stringent test on the quark model
dynamics~ \cite{landesberg,zhusl}, it is therefore highly desirable
to study the electromagnetic decay rates from excited baryon states
in order to refine the quark model description of the baryons. To
date very few electromagnetic transition rates have been measured
for the excited baryon resonances~\cite{PDG}. For a detailed
discussion of the experimental and theoretical status of the excited
baryons and their electromagnetic decays, see the review by
Landsberg~\cite{landesberg}.


\section{Estimation of branching ratios for $J/\psi \to B^*\bar B +
\bar B^* B \to B\bar{B}\gamma$ }

In hadron spectroscopy, the ground states of the hadron spectrum are
now well understood. However, the excited states still prove a
significant challenge. The first excited state $N^*(1440)P_{11}$
with positive parity $J^P=1/2^+$, and the adjacent excited state
$N^*(1535) S_{11}$ with negative parity $J^P=1/2^-$, as well as
$N^*(1520) D_{13}$ with $J^P=3/2^-$ have been identified by using
various techniques.  Although these four-star resonances are within
the energy region of many modern research facilities, their
properties including radiative decays are still not well determined.
Previous BES experiments already clearly observed these resonances
in $J/\psi\to p\bar p\eta$, $\bar pn\pi^++c.c.$, $\bar
pp\pi^0$~\cite{BES1,BES2}. With two orders of magnitude higher
statistics at BESIII, the radiative decays of these $N^*$ may also
be studied in $J/\psi\to \gamma p\bar p$. In fact, this decay
channel has already been studied by BESII experiment.  A strong
narrow peak $X(1860)$ near the threshold in the invariant mass
spectrum of proton-antiproton pairs was observed~\cite{BES}. The
branching ratio for $J/\psi \to \gamma p \bar p$ is about $3.8\times
10^{-4}$~\cite{PDG}, among which the contribution of $J/\psi \to
\gamma X(1860) \to \gamma p\bar p$ is about $7.0\times
10^{-5}$~\cite{BES}. The PWA formulae for determining quantum
numbers of intermediate resonances decaying to $p\bar p$ are given
in Ref.\cite{Dulat1}. Due to limited statistics and large background
from $J/\psi\to\bar pp\pi^0$ channel, no observation of $N^*\to
p\gamma$ was reported.

Based on the branching ratios for the reaction $J/\psi \to N^*\bar p
+ \bar N^* p$ measured by BESII~\cite{BES2} and branching ratios of
$N^*\to p\gamma$ given by PDG~\cite{PDG}, we give the estimation of
branching ratios for the reaction $J/\psi \to N^*\bar p + \bar N^* p
\to p\bar{p}\gamma$ through the intermediate $N^*=p(938)$,
$N^*(1440)$, $N^*(1535)$ and $N^*(1520)$ states as shown in
Table~\ref{threens}.

\begin{table}
\begin{center}
\caption{ The mass (MeV), widths (MeV), and branching ratios
($10^{-6}$) for $J/\psi \to N^* \bar p+\bar N^* p \to  p\gamma \bar
p$  through intermediate $N^*$ states.}\label{threens}
\begin{tabular}{ccccc}
\hline $M_{N^*}$  & $\Gamma$& $Br(J/\psi\!\to\! N^*\bar p\!+\!\bar N^*p)$ & $Br( N^* \to p\gamma)$ & $Br(J/\psi\!\to\! p\gamma\bar p)$    \\
\hline $938$      &         &   $210 \sim 224$\cite{PDG}           &                             & $19.8 \sim 21.0$                   \\
\hline $1440$     & $300$   &   $133 \sim 354$\cite{BES2}          & $350 \sim 480$  \cite{PDG}  & $0.046 \sim 0.170$                \\
\hline $1535$     & $150$   &   $ 92 \sim 210$\cite{BES2}          & $1500 \sim 3500$\cite{PDG}  & $0.138 \sim 0.735$                \\
\hline $1520$     & $115$   &   $ 34 \sim 154$\cite{BES2}          & $4600 \sim 5600$\cite{PDG}  & $0.156 \sim 0.862$                \\
\hline
\end{tabular}
\end{center}
\end{table}

In the estimation of the contribution from the off-shell nucleon
pole, we use the following effective lagrangian for the vertex
$\gamma pp$~\cite{Gao}
$$
{\mathcal {L}}_{\gamma pp}=-e\bar \psi_p(\gamma^\mu A_\mu
-\frac{\kappa_p}{2M_p}\sigma^{\mu\nu}\partial_\nu A_\mu)\psi_p\;,
$$
where $\kappa_p=2.739$ is the magnetic moment of the proton. The
following off-shell form factor is assumed
\begin{eqnarray}
F=\frac{\Lambda^4}{\Lambda^4+(p^2_{N^*}-m^2_{N^*})^2}\;,
\end{eqnarray}
with  $\Lambda=0.8GeV$. Here we also use the experimental photon
energy cut condition $ E_\gamma
>25MeV$. Because of the zero width of proton, the main contribution
for $J/\psi \to p\bar p \to p\bar{p}\gamma$ is from the low energy
photon, for example,  the branching ratio will be reduced to
$6.7\times10^{-6}$ for the photon energy cut $E_\gamma >100MeV$. The
contribution from the off-shell proton pole contribution is well
separated from those from $N^*$ contributions on the Dalitz plot.

Due to flavor SU(3) symmetry, the excited hyperons are produced at a
similar rate. So the typical branching ratio for the $J/\psi \to
B^*\bar B + \bar{B^*}B \to \gamma B\bar B$ processes is about
$10^{-7}\sim 10^{-6}$. With expected $10^{10}$ $J/\psi$ events and
much improved photon detection at BESIII, these
processes can definitely be studied in order to provide unique information on
the structure of various excited nucleon and hyperon states, and to give
 substantial insight into the non-perturbative aspects of the QCD.

\section{Formalism}

Now we present the necessary tools for the construction of covariant
L-S scheme for the $B^*\bar B M$($\bar B^*B M$) and
$B^*B\gamma$($\bar B^*\bar B\gamma$) couplings. The partial wave
amplitudes $U_i^{\mu\nu}$ in the covariant L-S scheme can be
constructed by using pure orbital angular momentum covariant tensors
$\tilde t^{(L_{bc})}_{\mu_1\cdots\mu_{L_{bc}}}$, covariant spin wave
functions  $\psi$ ($\Psi$) or $\phi$ ($\Phi$),  metric tensor
$g^{\mu\nu}$, totally antisymmetric Levi-Civita tensor
$\epsilon_{\mu\nu\lambda\sigma}$ and momentum of parent particle.

For a given hadronic decay process $a\to bc$, in the L-S scheme on
hadronic level, the initial state is described by its 4-momentum
$p_\mu$ and its spin state ${\bf S}_a$; the final state is described
by the relative orbital angular momentum state of $bc$ system $\bf
L$${_{bc}}$ and their spin states (${\bf S}_b$, ${\bf S}_c$). The
spin states (${\bf S}_a$, ${\bf S}_b$, ${\bf S}_c$) can be well
represented by the relativistic Rarita-Schwinger spin wave functions
for particles of arbitrary spin~\cite{Liang, Rarita,Chung,zhu}.  The
spin-$1\over 2$ wavefunction is the standard Dirac spinor $u(p,S)$
or $v(p,S)$ and the spin-1 wave function is the standard spin-1
polarization four-vector $\varepsilon^\mu(p,S)$ for a particle with
momentum p and spin projection $S$
\begin{equation}\label{spin one}
\sum_{S=0,\pm 1}\varepsilon_\mu(p,S)\varepsilon^*_\nu(p,S)
=-g_{\mu\nu}+{p_\mu p_\nu\over p^2}\equiv  -\tilde g_{\mu\nu}(p)
\end{equation}
with $p^\mu\varepsilon_\mu(p,S)  = 0$, which states that
spin-1 wave function is orthogonal to its own momentum. Here the
Minkowsky metric tensor has the form
$$
g_{\mu\nu} = diag(1,-1,-1,-1).
$$
Spin wave functions for particles of higher spins are constructed
from these two basic spin wave functions with C-G coefficients
$(J_1,J_{1z};J_2,J_{2z}|J,J_z)$ as
\begin{equation}
\label{epn}\varepsilon_{\mu_1\mu_2\cdots\mu_n}(p,n,S)
=\sum_{S_{n-1},S_n}(n-1,S_{n-1};1,S_n|n,S)
\varepsilon_{\mu_1\mu_2\cdots\mu_{n-1}}(p,n-1,S_{n-1})\varepsilon_{\mu_n}(p,S_n)
\end{equation}
for a particle with integer spin $n\ge 2$, and
\begin{equation}
\label{un}u_{\mu_1\mu_2\cdots\mu_n}(p,n+{1\over 2},S)
=\sum_{S_n,S_{n+1}}(n,S_n;{1\over 2},S_{n+1}|n+{1\over 2},S)
\varepsilon_{\mu_1\mu_2\cdots\mu_n}(p,n,S_n)u(p,S_{n+1})
\end{equation}
for a particle with half integer spin $n+{1\over 2}$ of $n\ge 1$.
For an antiparticle with half integer spin $n+{1\over 2}$ of $n\ge
1$, one has
\begin{equation}
v_{\mu_1\mu_2\cdots\mu_n}(p,n+{1\over 2},S)
=\sum_{S_n,S_{n+1}}(n,S_n;{1\over 2},S_{n+1}|n+{1\over 2},S)
\varepsilon_{\mu_1\mu_2\cdots\mu_n}(p,n,S_n)v(p,S_{n+1})
\end{equation}

For a process $a\to b + c$, if there exists a relative orbital
angular momentum ${\bf L}_{bc}$ between the particle b and c, then
the orbital angular momentum ${\bf L}_{bc}$ state can be represented
by covariant tensor wave functions $\tilde
t^{(L)}_{\mu_1\cdots\mu_L}$, which are the same as for meson
decay~\cite{Zou3,Chung,Andery}
\begin{eqnarray}
\tilde t^{(0)} &=& 1 , \\
\tilde t^{(1)}_\mu &=& \tilde g_{\mu\nu}(p_a)r^\nu\equiv\tilde r_\mu \;,\\
\tilde t^{(2)}_{\mu\nu} &=& \tilde r_\mu\tilde r_\nu
-{1\over 3}(\tilde r\cdot\tilde r)\tilde g_{\mu\nu}\;, \\
\tilde t^{(3)}_{\mu\nu\lambda} &=& \tilde r_\mu\tilde r_\nu\tilde
r_\lambda -{1\over 5}(\tilde r\cdot\tilde r)(\tilde g_{\mu\nu}\tilde
r_\lambda
+\tilde g_{\nu\lambda}\tilde r_\mu+\tilde g_{\lambda\mu}\tilde r_\nu)\;,\\
\tilde t^{(4)}_{\mu\nu\lambda\sigma} &=&\tilde r_\mu\tilde
r_\nu\tilde r_\lambda \tilde r_\sigma-{1\over 7}(\tilde
r\cdot\tilde r)(\tilde g_{\mu\nu}\tilde r_\lambda\tilde r_\sigma
+\tilde g_{\nu\lambda}\tilde r_\mu\tilde r_\sigma +\tilde
g_{\lambda\mu}\tilde r_\nu \tilde r_\sigma + \tilde
g_{\mu\sigma}\tilde r_\nu \tilde r_\lambda\nonumber\\ &+& \tilde
g_{\nu\sigma}\tilde r_\lambda \tilde r_\mu +
g_{\lambda\sigma}\tilde r_\mu \tilde r_\nu) +{1\over
35}(\tilde r\cdot\tilde r)^2(\tilde g_{\mu\nu} \tilde
g_{\lambda\sigma} + \tilde g_{\nu\lambda}\tilde g_{\mu\sigma}
+\tilde g_{\lambda\mu} \tilde
g_{\nu\sigma} )\;,\\
\tilde t^{(5)}_{\mu\nu\lambda\sigma\delta} &=&\tilde r_\mu\tilde
r_\nu\tilde r_\lambda \tilde r_\sigma \tilde r_\delta-{1\over
9}(\tilde r\cdot\tilde r)(\tilde g_{\mu\nu}\tilde r_\lambda\tilde
r_\sigma\tilde r_\delta +\tilde g_{\nu\lambda}\tilde r_\mu\tilde
r_\sigma\tilde r_\delta +\tilde g_{\lambda\mu}\tilde r_\nu \tilde
r_\sigma\tilde r_\delta + \tilde g_{\mu\sigma}\tilde r_\nu \tilde
r_\lambda\tilde r_\delta \nonumber\\
 &+& \tilde g_{\nu\sigma}\tilde r_\lambda
\tilde r_\mu\tilde r_\delta + g_{\lambda\sigma}\tilde r_\mu \tilde
r_\nu\tilde r_\delta + \tilde g_{\delta\mu}\tilde r_\lambda \tilde
r_\mu\tilde r_\sigma +\tilde g_{\delta\nu}\tilde r_\lambda \tilde
r_\mu\tilde r_\sigma + \tilde g_{\delta\sigma}\tilde r_\lambda
\tilde r_\mu\tilde r_\nu \nonumber\\&+&\tilde g_{\delta\lambda}\tilde r_\nu
\tilde r_\mu\tilde r_\sigma )
  +{1\over 63}(\tilde r\cdot\tilde
r)^2(\tilde g_{\mu\nu} \tilde g_{\lambda\sigma} \tilde r_\delta+
\tilde g_{\nu\lambda}\tilde g_{\mu\sigma} \tilde r_\delta +\tilde
g_{\lambda\mu} \tilde g_{\nu\sigma} \tilde r_\delta + \tilde
g_{\mu\nu} \tilde g_{\lambda\delta} \tilde r_\sigma\nonumber\\&+& \tilde
g_{\nu\lambda}\tilde g_{\mu\delta} \tilde r_\sigma +\tilde
g_{\lambda\mu} \tilde g_{\nu\delta} \tilde r_\sigma  + \tilde g_{\mu\nu} \tilde g_{\delta\sigma} \tilde r_\lambda+ \tilde
g_{\nu\delta}\tilde g_{\mu\sigma} \tilde r_\lambda +\tilde
g_{\delta\mu} \tilde g_{\nu\sigma} \tilde r_\lambda + \tilde
g_{\lambda\nu} \tilde g_{\delta\sigma} \tilde r_\mu\nonumber\\&+& \tilde
g_{\nu\delta}\tilde g_{\lambda\sigma} \tilde r_\mu +\tilde
g_{\delta\lambda} \tilde g_{\nu\sigma} \tilde r_\mu + \tilde
g_{\lambda\mu} \tilde g_{\delta\sigma} \tilde r_\nu+ \tilde
g_{\mu\delta}\tilde g_{\lambda\sigma} \tilde r_\nu +\tilde
g_{\delta\lambda} \tilde g_{\mu\sigma} \tilde r_\nu )\;,\\
 & & \cdots  \nonumber\\
\tilde t^{(L)}_{\mu_{i_1}\mu_{i_2}\cdots\mu_{i_L}}&=&\tilde
r_{\mu_{i_1}}\tilde r_{\mu_{i_1}}\cdots\tilde
r_{\mu_{i_L}}+\sum^{[L/2]}_{l=1}\sum^{L}_{i_1< i_2< \cdots <i_{2l}
=1}\frac{(-\tilde r\cdot\tilde
r)^l}{(2L-1)(2L-3)\cdots(2L-2l+1)}\nonumber\\
&&\times\frac{1}{2^{l}l!}(\tilde g_{\mu_{i_{1}}\mu_{i_{2}}}\tilde
g_{\mu_{i_{3}}\mu_{i_{4}}}\cdots \tilde
g_{\mu_{i_{2l-1}}\mu_{i_{2l}}}+ \mu_{i_1}, \mu_{i_2},\cdots
\mu_{i_{2l}}\; permutation, (2l)!\;  term)\nonumber\\
&&\times (\tilde r_{\mu_{1}}\tilde r_{\mu_{2}}\cdots\tilde
r_{\mu_{i_{1}-1}}\tilde r_{\mu_{i_{1}+1}}\cdots\tilde
r_{\mu_{i_{2}-1}}\tilde r_{\mu_{i_{2}+1}}\cdots\tilde
r_{\mu_{i_{2l}-1}}\tilde r_{\mu_{i_{2l}+1}}\cdots\tilde
r_{\mu_{L}}), \;
\end{eqnarray}
where  $r=p_b-p_c$ is the relative four momentum of the two decay
products in the parent particle rest frame; $(\tilde r\cdot\tilde
r)= -{\textbf{r}^2}$; $[L/2]=n$ when $L=2n$ and $L=2n+1$; and
$$
p_a^\mu \tilde t^{(1)}_\mu =p_a^\mu \tilde t^{(2)}_{\mu\nu} =p_a^\mu
t^{(3)}_{\mu\nu\lambda} =  0\;,\hspace{0.6cm}  \tilde
g^{\mu\nu}(p_a)= g^{\mu\nu}-{p_a^\mu p_a^\nu\over
p_a^2}\;,\hspace{0.6cm} g^{\mu\nu}=diag(1,-1,-1,-1)\;.
$$
In the L-S scheme, we need to use the conservation relation of total
angular momentum
\begin{equation}
\label{L} {\bf S}_a = {\bf S}_b+{\bf S}_c+{\bf L}_{bc}\quad or \quad
-{\bf S}_a + {\bf S}_b+{\bf S}_c+{\bf L}_{bc}=0.
\end{equation}
Besides the parity should be conserved, which means
\begin{equation}
\label{parity} \eta_a=\eta_b\eta_c(-1)^{L},
\end{equation}
where $\eta_a$, $\eta_b$ and $\eta_c$ are the intrinsic parities of
particles $a$, $b$ and $c$, respectively. From this relation, one
knows whether L should be even or odd. Then from Eq.(\ref{L}) one
can figure out how many different L-S combinations, which determine
the number of independent couplings.

Comparing with the pure meson case~\cite{Chung}, here we need to
introduce the concept of relativistic total spin of two fermions.
For the case of $a$ as a vector meson, $b$ as $B^*$ with spin
$n+{1\over 2}$ and $c$ as $\bar B$ with spin-$1\over 2$, the total
spin of bc (${\bf S}_{bc}$) can be either $n$ or $n+1$.  The two
${\bf S}_{bc}$ states can be represented as
\begin{eqnarray}
\psi^{(n)}_{\mu_1\cdots\mu_n} &=& \bar
u_{\mu_1\cdots\mu_n}(p_b,S_b)\gamma_5v(p_c,S_c),\\
\Psi^{(n+1)}_{\mu_1\cdots\mu_{n+1}} &=& \bar
u_{\mu_1\cdots\mu_n}(p_b,S_b)(\gamma_{\mu_{n+1}}-{r_{\mu_{n+1}}\over
m_a + m_b + m_c}) v(p_c,S_c)
\nonumber\\
& & +(\mu_1\leftrightarrow\mu_{n+1}) + \cdots +
(\mu_n\leftrightarrow\mu_{n+1})
\end{eqnarray}
for ${\bf S}_{bc}$ of $n$ and $n+1$, respectively. As a special case
of $n=0$, we have
\begin{eqnarray}
\psi^{(0)} &=& \bar u(p_b,S_b)\gamma_5v(p_c,S_c) , \\
\Psi^{(1)}_\mu &=& \bar u(p_b,S_b)(\gamma_\mu-{r_\mu\over m_a + m_b
+ m_c}) v(p_c,S_c) .
\end{eqnarray}
Here $r_\mu$ term is necessary to cancel out the $\bf\hat
p$-dependent component in the simple $\bar u\gamma_\mu v$
expression.

For the case of  $a$ as a vector meson, $b$ as excited anti-baryons
$(\bar B^*)$ with spin $n+{1\over 2}$ and c as baryons $( B)$ with
spin-$1\over 2$, the above equations can be written as
\begin{eqnarray}
 \psi^{C(n)}_{\mu_1\cdots\mu_n} &=& -\bar u(p_c,S_c)\gamma_5
v_{\mu_1\cdots\mu_n}(p_b,S_b),\\
\Psi^{C(n+1)}_{\mu_1\cdots\mu_{n+1}} &=& \bar
u(p_c,S_c)(\gamma_{\mu_{n+1}}-{r_{\mu_{n+1}}\over m_a + m_b + m_c})
v_{\mu_1\cdots\mu_n}(p_b,S_b)
\nonumber\\
& & +(\mu_1\leftrightarrow\mu_{n+1}) + \cdots +
(\mu_n\leftrightarrow\mu_{n+1})
\end{eqnarray}
for ${\bf S}_{bc}$ of $n$ and $n+1$, respectively. As a special case
of $n=0$, we have
\begin{eqnarray}
 \psi^{C(0)} &=& -\bar u(p_c,S_c)\gamma_5 v(p_b,S_b) , \\
\Psi^{C(1)}_\mu &=& \bar u(p_c,S_c)(\gamma_\mu-{r_\mu\over m_a + m_b
+ m_c}) v(p_b,S_b) .
\end{eqnarray}

For the case of $a$ as excited baryons ($B^*$) with spin $n+{1\over
2}$, $b$ as baryons ($B$) and $c$ as a meson, one needs to couple
$-{\bf S}_a$ and ${\bf S}_b$ first to get ${\bf S}_{ab}\equiv -{\bf
S}_a + {\bf S}_b$ states, which are
\begin{eqnarray}
\phi^{(n)}_{\mu_1\cdots\mu_n} &=& \bar
u(p_b,S_b)u_{\mu_1\cdots\mu_n}(p_a,S_a),\\
\Phi^{(n+1)}_{\mu_1\cdots\mu_{n+1}} &=& \bar
u(p_b,S_b)\gamma_5\tilde\gamma_{\mu_{n+1}}u_{\mu_1\cdots\mu_n}(p_a,S_a)
+ (\mu_1\leftrightarrow\mu_{n+1}) + \cdots +
(\mu_n\leftrightarrow\mu_{n+1})
\nonumber\\
\end{eqnarray}
for ${\bf S}_{ab}$ of $n$ and $n+1$, respectively.
\begin{eqnarray}
\phi^{(0)} &=& \bar u(p_b,S_b)u(p_a,S_a) , \\
\Phi^{(1)}_\mu &=& \bar u(p_b,S_b)\gamma_5\tilde\gamma_\mu
u(p_a,S_a)
\end{eqnarray}
with $\tilde\gamma_\mu=\tilde g_{\mu\nu}(p_a)\gamma^\nu$.

For the case of $a$ as excited antibaryons ($\bar B^*$) with spin
$n+{1\over 2}$, $b$ as an antibaryon ($\bar B$) and $c$ as a meson,
as before one needs to couple $-{\bf S}_a$ and ${\bf S}_b$ first to
get ${\bf S}_{ab}\equiv - {\bf S}_a+ {\bf S}_b$ states, which are
\begin{eqnarray}
\phi^{C(n)}_{\mu_1\cdots\mu_n} &=& \bar
v_{\mu_1\cdots\mu_n}(p_a,S_a)v(p_b,S_b),\\
\Phi^{C(n+1)}_{\mu_1\cdots\mu_{n+1}} &=& \bar
v_{\mu_1\cdots\mu_n}(p_a,S_a)\gamma_5\tilde\gamma_{\mu_{n+1}}
v(p_b,S_b) + (\mu_1\leftrightarrow\mu_{n+1}) + \cdots +
(\mu_n\leftrightarrow\mu_{n+1})
\nonumber\\
\end{eqnarray}
for ${\bf S}_{ab}$ of $n$ and $n+1$, respectively. As a special case
of $n=0$, we have
\begin{eqnarray}
\phi^{C(0)} &=& \bar v(p_a,S_a) v(p_b,S_b) , \\
\Phi^{C(1)}_\mu &=& \bar v(p_a,S_a)\gamma_5\tilde\gamma_\mu
v(p_b,S_b).
\end{eqnarray}

\section{Partial Wave Amplitudes}

We consider the following process
\begin{equation}\label{process-1}
J/\psi \to B^*\bar B + \bar B^* B \to \gamma  B \bar B\;,
\end{equation}
The possible $J^P$ for $B^*$ is ${1\over 2}^{\pm}$, ${3\over
2}^{\pm}$, ${5\over 2}^{\pm}$,  ${7\over 2}^{\pm}$. We denote the
four momenta of $J/\psi$, $B^*(\bar B^*)$ and $\gamma$ by $p_{\mu}$,
$p_{B^* \mu}(p_{\bar B^*\mu})$ and $q_\mu$ . The orbital spin tensor
describing the first and second steps will be denoted by $\tilde
T^{(L)}_{\mu_1\cdots\mu_L}$ and $\tilde
t^{(l)}_{\mu_1\cdots\mu_{l}}$. For the process (\ref{process-1}) the
general form of the decay amplitude is
\begin{equation}
M=\varepsilon_\mu(p,S_{J/\psi})e^*_\nu(q,S_{\gamma})M^{\mu\nu}
=\varepsilon_\mu(p,S_{J/\psi})e^*_\nu(q,S_{\gamma})\sum_{i,j}
U^{\mu\nu}_{i,j} \;,
\end{equation}
where $\varepsilon_\mu(p,S_{J/\psi})$ is the polarization four
vector of the $J/\psi$; $e_\nu(q,S_\gamma)$ is the polarization four
vector of the photon; ${\bf S}_{J/\psi}$ and $S_\gamma$ are the
spins of $J/\psi$ and photon; $U_{i,j}^{\mu\nu}$ is the $i$-th $B^*$
and $\bar B^*$, j-th partial wave amplitude with complex coupling
constants to be determined by the experiment. The spin-1
polarization four vector $\varepsilon_\mu(p,S_{J/\psi})$ for
$J/\psi$ with four momentum $p_\mu$ satisfies the relation in
Eq.(\ref{spin one}). For $J/\psi$ production from $e^+ e^-$
annihilation, the electrons are highly relativistic, with the result
that $J_z = \pm 1$ for the $J/\psi$ spin projection taking the beam
direction as the z-axis. This limits $S_{J/\psi}$ to 1 and 2. Then one has the following relation
\begin{equation}\label{jpsi}
\sum^2_{S_{J/\psi}=1}\varepsilon_{\mu}(p,S_{J/\psi})\varepsilon^*_{\mu'}(p,S_{J/\psi})
=\delta_{\mu\mu'}(\delta_{\mu 1}+\delta_{\mu 2}).
\end{equation}
For the  photon polarization four vector, there is the usual Lorentz
orthogonality conditions. Namely, the polarization four vector
$e_\nu(q,S_\gamma)$ of the  photon with  momenta $q$ satisfies
\begin{equation}
q^\nu e_\nu(q,S_\gamma)=0,
\end{equation}
which states that spin-1 wave function is orthogonal to its own
momentum. The above relation is the same as for a massive vector
meson. However, for the photon, there is an additional gauge
invariance condition
\begin{equation}\label{photon}
\sum_{S_\gamma}e^*_\mu(q,S_\gamma) e_\nu(q,S_\gamma)=
-g_{\mu\nu}+\frac{q_\mu K_\nu+ K_\mu q_\nu}{q\!\cdot\!
K}-\frac{K\!\cdot\! K}{(q\!\cdot\! K)^2}q_\mu q_\nu \equiv
-g^{(\perp\perp)}_{\mu\nu}
\end{equation}
with $K=p-q$ and $K^\nu e_\nu =0$.

Although $B^*({1\over 2}^{\pm}$, ${3\over 2}^{\pm}$, ${5\over
2}^{\pm}$, ${7\over 2}^{\pm}) B\omega$ couplings have the same
structure as the $B^*({1\over 2}^{\pm}$, ${3\over 2}^{\pm}$,
${5\over 2}^{\pm}$, ${7\over 2}^{\pm})B\gamma$ couplings, the gauge
invariance requirement for the $B^* B\gamma$ couplings reduces the
number of independent amplitudes. For example, the partial wave
amplitudes for the process $B^*({3\over2}^+) \to
B({1\over2}^+)\gamma$ can be written as
\begin{equation}
M = (g_1M^\nu_1+g_2M^\nu_2+g_3M^\nu_3)e^*_\nu(q,S_\gamma)\;,\label{amp}\\
\end{equation}
where
\begin{equation}
M^\nu_1=i \phi^{(1)}_{\mu}\epsilon^{\mu\nu\lambda\sigma}
\tilde{t}^{(1)}_{\lambda}\hat p_{B^*\sigma}\;,\ \ \ \
M^\nu_2=\Phi^{(2)\nu\mu}\tilde{t}^{(1)}_{\mu}\;,\ \ \ \
M^\nu_3=\Phi^{(2)}_{\mu\lambda}\tilde{t}^{(3)\mu\lambda\nu}\;.
\end{equation}
Because of the gauge invariance requirement
\begin{equation}
(g_1M^\nu_1+g_2M^\nu_2+g_3M^\nu_3)q_\nu=0\;,\label{amp}\\
\end{equation}
we get the following relation
\begin{equation}
g_2=-\frac{3}{5}(\tilde{r}\cdot\tilde{r})\;g_3=\frac{3}{5}\frac{(m^2_{B^*}-m^2_{B})^2}{m^2_{B^*}}\;g_3\label{gauge}\;,
\end{equation}
which means that there are  two independent partial wave amplitudes.
Refs.\cite{Anisovich_1,Anisovich_2} also provided basically
equivalent partial wave amplitude formulae for the vertex
$B^*B\gamma$ in the spin-orbital approach. In order to be able to
compare our results with conventional helicity amplitudes for the
radiative decays of baryon resonances~\cite{PDG}, we also give the
relation between our coupling constants and helicity amplitudes in
the Appendix.

To compute decay width, we need an expression for $|M|^2$. Note that
the square modulus of the decay amplitude, which gives the decay
probability of a certain configuration should be independent of any
particular frame, that is, a Lorentz scalar. Thus by using the Eqs.
(\ref{jpsi}) and (\ref{photon}), we have
\begin{equation}
d\Gamma = \frac{(2\pi)^4}{2M_{J/\psi}}|\textit{M}|^2d\Phi_3(p;
q_\gamma, p_B, p_{\bar B})\;,
\end{equation}
where $M_{J/\psi}$ is the mass of the $J/\psi$, and the general form
of the matrix element square is
\begin{eqnarray}
|\textit{M}|^2 &=& \frac{1}{2}
\sum^2_{S_{J/\psi}=1}\sum^2_{S_\gamma=1} \sum_{S_B} \sum_{S_{\bar
B}} |\varepsilon_{\mu}(p,S_{J/\psi})
e^*_{\nu}(q,S_\gamma)M^{\mu\nu}|^2 \nonumber\\ &=&
 \frac{1}{2}\sum^2_{\mu=1} \sum_{S_p} \sum_{S_{\bar
p}} M^{\mu\nu} (-g^{(\perp\perp)}_{\nu\nu'}) M^{*\mu\nu'}
\nonumber\\
&=&  \frac{1}{2}\sum_{i,i^\prime}\sum_{j,j^\prime}\sum^2_{\mu=1}
\sum_{S_p} \sum_{S_{\bar p}}
U_{i,j}^{\mu\nu}(-g^{(\perp\perp)}_{\nu\nu'})U_{i^\prime,j^\prime}^{*\mu\nu'}\;,
\end{eqnarray}
the standard Lorentz invariant 3-body phase space element $d\Phi_3$
is given by
\begin{equation}
d\Phi_3(p; q, p_B, p_{\bar B})=\delta^4(p - q - p_B - p_{\bar B})
\frac{d^3{\bf q}}{(2\pi)^3 2E_\gamma }  \frac{2m_B d^3{\bf
p}_B}{(2\pi)^3 2E_B } \frac{2 m_{\bar B} d^3{\bf p}_{\bar B} }
{(2\pi)^3 2E_{\bar B} }\;.
\end{equation}
From (\ref{process-1}) we see that $B^*$ and $\bar B^*$ are the
intermediate resonances decaying into $B\gamma$ and $\bar B^*\gamma$
respectively, therefore we need to introduce into the amplitude  the
following propagators denoted by $G_{B^*}$ and $G_{\bar B^*}$
\cite{ouyang, zou-pwa}
\begin{eqnarray}\label{propagator}
G_{B^*}(\frac{1}{2}) &=& f^{B^*}_{B\gamma} \sum_{S_{B^*}}
u(p_{B^*},S_{B^*})\bar u(p_{B^*},S_{B^*})
=f^{B^*}_{B\gamma}\frac{(\not\! p_{B^*}+
m_{B^*)}}{2m_{B^*}},\nonumber\\
G_{\bar B^*}(\frac{1}{2})&=&\bar f^{\bar B^*}_{\bar B\gamma}
\sum_{S_{\bar B^*}} v(p_{\bar B^*},S_{\bar B^*})\bar v(p_{\bar
B^*},S_{\bar B^*}) =\bar f^{\bar B^*}_{\bar B\gamma} \frac{(\not\!
p_{\bar B^*} - m_{\bar B^*)}}{2m_{\bar B^*}},\\
G^{\mu\nu}_{B^*}(\frac{3}{2}) &=& f^{B^*}_{B\gamma}
\sum_{S_{B^*}}u^{\mu}(p_{B^*},S_{B^*})\bar u^{\nu}(p_{B^*},S_{B^*})=
f^{B^*}_{B\gamma}\frac{(\not\! p_{B^*} + m_{B^*}
)}{2m_{B^*}}P^{\mu\nu}_{B^*}(\frac{3}{2}),\nonumber\\
G^{\mu\nu}_{\bar B^*}(\frac{3}{2}) &=& \bar f^{\bar B^*}_{\bar
B\gamma}\sum_{S_{\bar B^*}}v^{\mu}(p_{\bar B^*},S_{\bar B^*})\bar
v^{\nu}(p_{\bar B^*},S_{\bar B^*})=\bar f^{\bar B^*}_{\bar
B\gamma}\frac{(\not\! p_{\bar B^*} - m_{\bar B^*} )}{2m_{\bar
B^*}}P^{\mu\nu}_{\bar B^*}(\frac{3}{2}),\\
G^{\mu\nu\alpha\beta}_{B^*}(\frac{5}{2})&=&f^{B^*}_{B\gamma}
\sum_{S_{B^*}}u^{\mu\nu}(p_{B^*},S_{B^*})\bar
u^{\alpha\beta}(p_{B^*},S_{B^*})= f^{B^*}_{B\gamma}\frac{(\not\!
p_{B^*} + m_{B^*}
)}{2m_{B^*}}P^{\mu\nu}_{B^*}(\frac{5}{2}),\nonumber\\
G^{\mu\nu\alpha\beta}_{\bar B^*}(\frac{5}{2}) &=& \bar f^{\bar
B^*}_{\bar B\gamma}\sum_{S_{\bar B^*}}v^{\mu\nu}(p_{\bar
B^*},S_{\bar B^*})\bar v^{\alpha\beta}(p_{\bar B^*},S_{\bar
B^*})=\bar f^{\bar B^*}_{\bar B\gamma}\frac{(\not\! p_{\bar B^*} -
m_{\bar B^*} )}{2m_{\bar
B^*}}P^{\mu\nu}_{\bar B^*}(\frac{5}{2}),\\
G^{\mu\nu\alpha\beta\lambda\sigma}_{B^*}(\frac{7}{2}) &=& f^{
B^*}_{B\gamma}\sum_{S_{B^*}}u^{\mu\nu\alpha}(p_{B^*},S_{ B^*})\bar
u^{\beta\lambda\sigma}(p_{B^*},S_{B^*})= f^{B^*}_{\bar
B\gamma}\frac{(\not\! p_{B^*} + m_{B^*} )}{2m_{
B^*}}P^{\mu\nu\alpha\beta\lambda\sigma}_{B^*}(\frac{7}{2}),\nonumber\\
G^{\mu\nu\alpha\beta\lambda\sigma}_{\bar B^*}(\frac{7}{2}) &=& \bar
f^{\bar B^*}_{\bar B\gamma}\sum_{S_{\bar
B^*}}v^{\mu\nu\alpha}(p_{\bar B^*},S_{\bar B^*})\bar
v^{\beta\lambda\sigma}(p_{\bar B^*},S_{\bar B^*})= \bar f^{\bar
B^*}_{\bar B\gamma}\frac{(\not\! p_{\bar B^*} - m_{\bar B^*}
)}{2m_{\bar B^*}}P^{\mu\nu\alpha\beta\lambda\sigma}_{\bar
B^*}(\frac{7}{2}),\nonumber\\
\end{eqnarray}
where
\begin{equation}\label{denomenator}
f^{B^*}_{B\gamma}= \frac{2m_{B^*}}{p^2_{B^*} - m^2_{B^*} +
im_{B^*}\Gamma_{B^*}},\ \ \ \ \
 \bar f^{\bar B^*}_{\bar B\gamma}=
\frac{2m_{\bar B^*}}{p^2_{\bar B^*} - m^2_{\bar B^*} + im_{\bar
B^*}\Gamma_{\bar B^*}},
\end{equation}
here $m_{B^*}$, $m_{\bar B^*}$ and $\Gamma_{B^*}$, $\Gamma_{\bar
B^*}$ are the resonances masses and widths;
\begin{eqnarray}
P^{\mu\nu}_{B^*}(\frac{3}{2})&=&-g^{\mu\nu}+\frac{1}{3}\gamma^{\mu}\gamma^{\nu}
+ \frac{2}{3}\frac{p_{B^*}^{\mu}p_{B^*}^{\nu}}{m_{B^*}^{2}}+
\frac{1}{3m_{B^*}}(\gamma^{\mu}p_{B^*}^{\nu}-\gamma^{\nu}p_{B^*}^{\mu}),\nonumber\\
P^{\mu\nu}_{\bar B^*} (\frac{3}{2})&=&
-g^{\mu\nu}+\frac{1}{3}\gamma^{\mu}\gamma^{\nu} + \frac{2}{3}\frac{
p_{\bar B^*}^{\mu} p_{\bar B^*}^{\nu}}{m_{\bar B^*}^{2}}-
\frac{1}{3m_{\bar B^*}}(\gamma^{\mu}p_{\bar
B^*}^{\nu}-\gamma^{\nu}p_{\bar B^*}^{\mu}),\nonumber\\
P^{\mu\nu\alpha\beta}_{ B^*/\bar B^*}(\frac{5}{2}) &=&
\frac{1}{2}(\tilde g^{\mu\nu}\tilde g^{\alpha\beta} + \tilde
g^{\mu\beta}\tilde g^{\alpha\nu}) -\frac{1}{5}\tilde
g^{\mu\alpha}\tilde g^{\nu\beta}\nonumber\\
&-&\frac{1}{10}(\tilde\gamma^\mu\tilde\gamma^\nu \tilde
g^{\alpha\beta}+\tilde\gamma^\mu \tilde\gamma^\beta\tilde
g^{\alpha\nu}  +\tilde\gamma^\alpha\tilde\nu\tilde g^{\mu\beta} +
\tilde\gamma^\alpha\tilde\gamma^\beta\tilde g^{\mu\nu}),\\
P^{\mu\nu\alpha\beta\lambda\sigma}_{ B^*/\bar B^*}(\frac{7}{2}) &=&
\frac{4}{9}\gamma_\tau\gamma_\rho
P^{(4)\tau\mu\nu\alpha\rho\beta\lambda\sigma},\nonumber
\end{eqnarray}
and where
\begin{eqnarray}
P^{(4)\tau\mu\nu\alpha\rho\beta\lambda\sigma} &=& \frac{1}{24}\left(
\tilde g^{\tau\rho}\tilde g^{\mu\beta}\tilde g^{\nu\lambda}\tilde
g^{\alpha\sigma} +(\rho,\beta,\lambda,\sigma \;\; permutation,\;\;
24\;\; terms)\right) \nonumber\\
&-&\frac{1}{84}(  \tilde g^{\tau\mu}\tilde g^{\rho\beta}\tilde
g^{\nu\lambda}\tilde g^{\alpha\sigma} + (\tau,\mu, \nu,\alpha \;\;
permutation,\;\; \rho,\beta,\lambda,\sigma \;\;
\nonumber\\
&&\hspace{5cm}permutation, \;\; 72\;\; terms) \nonumber\\
&+& \frac{1}{105}(\tilde g^{\tau\mu} \tilde g^{\nu\alpha}+ \tilde
g^{\tau\nu}\tilde g^{\mu\alpha}+\tilde g^{\tau\alpha} \tilde
g^{\mu\nu} )( \tilde g^{\rho\beta}\tilde g^{\lambda\sigma}+ \tilde
g^{\rho\lambda}\tilde g^{\beta\sigma} + \tilde g^{\rho\sigma} \tilde
g^{\beta\lambda}) .\nonumber\\
\end{eqnarray}

For the different partial wave amplitudes, we use the following
notation
\begin{eqnarray}
(S_{B^*\bar{B}/\bar{B}^*B}\;,\; L_{B^*\bar{B}/\bar{B}^*B}\;,\;
S_{B^*B/\bar{B}^*\bar{B}}),\nonumber
\end{eqnarray}
where  ${\bf S}_{B^*\bar{B}/\bar{B}^*B}={\bf S}_{B^*}+{\bf
S}_{\bar{B}}$ or ${\bf S}_{\bar B^*}+{\bf S}_{B}$; ${\bf
L}_{B^*\bar{B}/\bar{B}^*B}$ is the relative orbital angular momentum
between $B^*$ and $\bar{B}$ or $\bar B^*$ and $B$; ${\bf
S}_{B^*B/\bar{B}^*\bar{B}}=-{\bf S}_{B^*}+{\bf S}_{B}$ or $-{\bf
S}_{\bar B^*}+{\bf S}_{\bar B}$. In the following by considering the
parity and angular momentum conservations  we provide all relevant
covariant amplitudes for the process (\ref{process-1}). In these
amplitudes, $1$, $2$ and $3$ denote the
 three final state particles $B$, $\bar{B}$ and $\gamma$.

For $J/\psi (1^-) \to B^*({1\over 2}^+) \bar B({1\over 2}^-) +
\bar B^*({1\over 2}^-) B({1\over 2}^+)\to \gamma B({1\over 2}^+)
\bar B({1\over 2}^-)$, we find two independent covariant
amplitudes for a vector meson $J/\psi (1^-)$ decaying into the
$B^*({1\over 2}^+) \bar B({1\over 2}^-)$ and $\bar B^*({1\over
2}^-) B({1\over 2}^+)$ states, and one independent covariant
amplitudes for a excited baryon resonances $B^*({1\over 2}^+)$ and
$\bar B^*({1\over 2}^-)$ decaying into $\gamma  B({1\over 2}^+)$
and $\gamma \bar B({1\over 2}^-)$. All in all we get the following
two covariant amplitudes with two independent coupling constants
$g^{i,a}$ and $g^{i,b}$ which are determined by the experiment
\begin{eqnarray}
(1,0,1)\nonumber\\
 U^{\mu\nu}_{i,1} &=&  g^{i,a}\;
 (\; \sum_{S_{B^*}}i\Phi^{(1)}_{\beta}
 \Psi^{(1)\mu}
 \tilde t^{(1)}_{(13)\lambda}\epsilon^{\beta\nu\lambda\sigma}
 \hat p_{ B^*\sigma}f^{B^*}_{B\gamma} \nonumber\\
 &&\ \ \ - \;\sum_{S_{\bar B^*}}i\Psi^{C(1)\mu}\Phi^{C(1)}_{\beta} \tilde
t^{(1)}_{(23)\lambda}\epsilon^{\beta\nu\lambda\sigma}
\hat p_{\bar B^*\sigma}\bar f^{\bar B^*}_{\bar B \gamma}\;)\;,\\
(1,2,1)\nonumber\\
 U^{\mu\nu}_{i,2} &=&  g^{i,b}\; (\;\sum_{S_{B^*}}
  i\Phi^{(1)}_{\beta}\Psi^{(1)}_{\alpha}\;
 \tilde T^{(2)\alpha\mu}_{(B^*2)}\tilde t^{(1)}_{(13)\lambda}
 \epsilon^{\beta\nu\lambda\sigma}
 \hat p_{B^* \sigma}f^{B^*}_{B\gamma}\nonumber\\
 &&\ \ \ - \;\sum_{S_{\bar B^*}}
i\Psi^{C(1)}_{\alpha}\Phi^{C(1)}_{\beta}
  \tilde T^{(2)\alpha\mu}_{(\bar B^*1)}
  \tilde t^{(1)}_{(23)\lambda}\epsilon^{\beta\nu\lambda\sigma}
  \hat p_{\bar B^* \sigma}\bar f^{\bar B^*}_{\bar B \gamma}\; ) \;,
\end{eqnarray}

For $J/\psi (1^-) \to B^*({1\over 2}^-) \bar B({1\over 2}^-) +
\bar B^*({1\over 2}^+) B({1\over 2}^+)\to \gamma B({1\over 2}^+)
\bar B({1\over 2}^-)$ we get the following two covariant
amplitudes with two independent coupling constants $g^{i,a}$ and
$g^{i,b}$ which are determined by the experiment
\begin{eqnarray}
(0,1,1) \nonumber\\
U^{\mu\nu}_{i,1}
&=&g^{i,a}\;(\;-\;\frac{2}{3}C_{B^*B\gamma}\sum_{S_{B^*}}\Phi^{(1)\nu}
\psi^{(0)}\tilde
T^{(1)\mu}_{(B^*2)}f^{B^*}_{B\gamma} \nonumber \\
 &&\ \ \ \ \ \ \ -\;\frac{2}{3}C_{\bar{B}^*\bar{B}\gamma} \sum_{S_{\bar B^*}}
 \psi^{C(0)}\Phi^{C(1)\nu}\tilde T^{(1)\mu}_{(\bar B^* 1)}
\bar f^{\bar B^*}_{\bar B \gamma}\;\nonumber\\
 &&\ \ \ \ \ \ \ + \;\sum_{S_{B^*}}
\Phi^{(1)}_{\beta}\psi^{(0)}\tilde T^{(1)\mu}_{(B^*2)} \tilde
t^{(2)\beta\nu}_{(13)}f^{B^*}_{B\gamma}\;\nonumber\\
 &&\ \ \ \ \ \ \ + \;\sum_{S_{\bar B^*}}\psi^{C(0)}\tilde
\Phi^{C(1)}_{\beta}T^{(1)\mu}_{(\bar B^* 1)}\tilde
t^{(2)\beta\nu}_{(23)}\bar f^{\bar B^*}_{\bar B \gamma}\ \ ),
\end{eqnarray}
\begin{eqnarray}
(1,1,1) \nonumber\\
 U^{\mu\nu}_{i,2} &=& g^{i,b}\;( -\;\frac{2}{3}C_{B^*B\gamma}
 \sum_{S_{B^*}}i \Phi^{(1)\nu} \Psi^{(1)}_{\alpha}
 \tilde T^{(1)}_{(B^*2)\delta}\epsilon^{\alpha\mu\delta\tau}
 \hat p_{\tau}f^{B^*}_{B\gamma} \nonumber\\
 &&\ \ \ \ \ \ \ +\; \frac{2}{3}C_{\bar{B}^*\bar{B}\gamma}\;\sum_{S_{\bar B^*}}
 i\Psi^{C(1)}_{\alpha}\Phi^{C(1)\nu}
 \tilde T^{(1)}_{(\bar B^* 1)\delta}
 \epsilon^{\alpha\mu\delta\tau}\hat p_{\tau}
 \bar f^{\bar B^*}_{\bar B \gamma} \;)\nonumber\\
&&\ \ \ \ \ \ \ +\;\sum_{S_{B^*}}i
\Phi^{(1)}_{\beta}\Psi^{(1)}_{\alpha}
 \tilde T^{(1)}_{(B^*2)\delta}\tilde t^{(2)\beta\nu}_{(13)}
  \epsilon^{\alpha\mu\delta\tau}\hat p_{\tau}
 f^{B^*}_{B\gamma}\nonumber\\
 &&\ \ \ \ \ \ \ - \;\sum_{S_{\bar B^*}}
 i\Psi^{C(1)}_{\alpha}\Phi^{C(1)}_{\beta}
 \tilde T^{(1)}_{(\bar B^* 1)\delta}
 \tilde t^{(2)\beta\nu}_{(23)}\epsilon^{\alpha\mu\delta\tau} \hat p_{\tau}
  \bar f^{\bar B^*}_{\bar B \gamma}.
\end{eqnarray}

One may note that for $J/\psi \to B^*\bar B + \bar B^* B \to \gamma
B \bar B$ with $J^P(B^*(\bar{B}^*))={3\over 2}^{\pm}$, ${5\over
2}^{\pm}$, ${7\over 2}^{\pm}$, we get the six covariant amplitudes
for i-th $B^*(\bar{B}^*)$ with five independent coupling constants
$g^{i,a}_{J/\psi}$, $g^{i,b}_{J/\psi}$, $g^{i,c}_{J/\psi}$,
$g^{i,a}_{\gamma}$ and $g^{i,b}_{\gamma}$ which are determined by
the experiment. Thus for $J/\psi (1^-) \to B^*({3\over 2}^+) \bar
B({1\over 2}^-) + \bar B^*({3\over 2}^-)  B({1\over 2}^+) \to \gamma
B({1\over 2}^+) \bar B({1\over 2}^-)$ we get the following six
covariant amplitudes
\begin{eqnarray}
(1,0,1)\nonumber\\
 U^{\mu\nu}_{i,1} &=& \;g^{i,a}_{J/\psi}\;g^{i,a}_{\gamma}\; (\;\sum_{S_{B^*}}
 i\phi^{(1)}_{\beta}\psi^{(1)\mu}
 \tilde t^{(1)}_{(13)\lambda}\epsilon^{\beta\nu\lambda\sigma}
 \hat p_{B^*\sigma}f^{B^*}_{B\gamma}\nonumber\\
 &&\ \ \ \ \ \ \ \ \ \ \ - \;\sum_{S_{\bar B^*}}
 i\psi^{C(1)\mu} \phi^{C(1)}_{\beta}
 \tilde t^{(1)}_{(23)\lambda}\epsilon^{\beta\nu\lambda\sigma}
 \hat p_{\bar B^*\sigma}
 \bar f^{\bar B^*}_{\bar B \gamma}\;)\;,\\ 
(1,2,1)\nonumber\\
 U^{\mu\nu}_{i,2} &=&  \;g^{i,b}_{J/\psi}\;g^{i,a}_{\gamma}\;(\;\sum_{S_{B^*}}
  i\phi^{(1)}_{\beta}\psi^{(1)}_{\alpha}\tilde T^{(2)\alpha\mu}_{(B^*2)}
 \tilde t^{(1)}_{(13)\lambda} \epsilon^{\beta\nu\lambda\sigma}
\hat p_{B^*\sigma}f^{B^*}_{B\gamma}\nonumber\\
 &&\ \ \ \ \ \ \ \ \ \ \ -\;\sum_{S_{\bar B^*}} i\psi^{C(1)}_{\alpha}
 \phi^{C(1)}_{\beta}
 \tilde T^{(2)\alpha\mu}_{(\bar B^*1)}\tilde t^{(1)}_{(23)\lambda}
 \epsilon^{\beta\nu\lambda\sigma}
\hat p_{\bar B^*\sigma}
 \bar f^{\bar B^*}_{\bar B \gamma}\;)\;,\\
(2,2,1)\nonumber\\
U^{\mu\nu}_{i,3} &=& g^{i,c}_{J/\psi}\;g^{i,a}_{\gamma}\;
(\;-\sum_{S_{B^*}}
\phi^{(1)}_{\beta}\Psi^{(2)}_{\alpha\rho}\epsilon^{\alpha\mu\delta\tau}\tilde
T^{(2)\rho}_{(B^*2)\delta}\hat
p_{\tau}\epsilon^{\beta\nu\lambda\sigma}\tilde
t^{(1)}_{(13)\lambda}\hat p_{B^*\sigma}f^{B^*}_{B\gamma}\nonumber\\
 &&\ \ \ \ \ \ \ \ \ \ \ \ - \; \sum_{S_{\bar B^*}}
\Psi^{C(2)}_{\alpha\rho}\phi^{C(1)}_{\beta}\epsilon^{\alpha\mu\delta\tau}\tilde
T^{C(2)\rho}_{(\bar B^* 1)\delta}\hat
p_{\tau}\epsilon^{\beta\nu\lambda\sigma}\tilde
t^{C(1)}_{(23)\lambda}\hat p_{\bar B^*\sigma}\bar f^{\bar B^*}_{\bar B \gamma}\; )\;,\\
(1,0,2)\nonumber\\
U^{\mu\nu}_{i,4} &=& g^{i,a}_{J/\psi}\;g^{i,b}_{\gamma}\;
(-\;\frac{3}{5}C_{B^*B\gamma}\;\sum_{S_{B^*}}\Phi^{(2)\nu\beta}
\psi^{(1)\mu}\tilde
t^{(1)}_{(13)\beta}f^{B^*}_{B\gamma}\nonumber\\
 &&\ \ \ \ \ \ \ \ \ \ \ - \;\frac{3}{5}C_{\bar{B}^*\bar{B}\gamma}\; \sum_{S_{\bar B^*}}
 \psi^{C(1)\mu}\Phi^{C(2)\nu\beta}\tilde t^{C(1)}_{(23)\beta}\bar f^{\bar B^*}_{\bar B \gamma}\;
\;\nonumber\\
 &&\ \ \ \ \ \ \ \ \ \ \ +\;\sum_{S_{B^*}}\Phi^{(2)}_{\beta\lambda}\psi^{(1)\mu}
\tilde t^{(3)\beta\lambda\nu}_{(13)}f^{B^*}_{B\gamma}\nonumber\\
 &&\ \ \ \ \ \ \ \ \ \ \ + \; \sum_{S_{\bar B^*}}
\psi^{C(1)\mu} \Phi^{C(2)}_{\beta\lambda}\tilde
t^{(3)\beta\lambda\nu}_{(23)}\bar f^{\bar B^*}_{\bar B \gamma}\;\ \
\ \  )
\;,\\
(1,2,2)\nonumber\\
U^{\mu\nu}_{i,5} &=& g^{i,b}_{J/\psi}\;g^{i,b}_{\gamma}\;
(-\;\frac{3}{5}C_{B^*B\gamma}\;\sum_{S_{B^*}}\tilde
\Phi^{(2)\nu\beta}
\psi^{(1)}_{\alpha}T^{(2)\alpha\mu}_{(B^*2)}\tilde
t^{(1)}_{(13)\beta}f^{B^*}_{B\gamma}\nonumber\\
 &&\ \ \ \ \ \ \ \ \ \ \  -\; \frac{3}{5}C_{\bar{B}^*\bar{B}\gamma}\; \sum_{S_{\bar B^*}}
\psi^{C(1)}_{\alpha}\Phi^{C(2)\nu\beta}\tilde
T^{(2)\alpha\mu}_{(\bar B^*1)}\tilde
t^{(1)}_{(23)\beta}\bar f^{\bar B^*}_{\bar B \gamma}\;)\;\nonumber\\
 &&\ \ \ \ \ \ \ \ \ \ \ + \;\sum_{S_{B^*}}\Phi^{(2)}_{\beta\lambda}
\psi^{(1)}_{\alpha}\tilde T^{(2)\alpha\mu}_{(B^*2)}
\tilde t^{(3)\beta\lambda\nu}_{(13)}f^{B^*}_{B\gamma}\nonumber\\
 &&\ \ \ \ \ \ \ \ \ \ \ + \; \sum_{S_{\bar B^*}}
\psi^{C(1)}_{\alpha} \Phi^{C(2)}_{\beta\lambda}\tilde
T^{(2)\alpha\mu}_{(\bar B^*1)}\tilde t^{(3)\beta\lambda\nu}_{(23)}\
\bar f^{\bar B^*}_{\bar B \gamma}\;)\;,\\
(2,2,2)\nonumber\\
U^{\mu\nu}_{i,6} &=&
g^{i,c}_{J/\psi}\;g^{i,b}_{\gamma}(-\;\frac{3}{5}C_{B^*B\gamma}\sum_{S_{B^*}}i
\Phi^{(2)\nu\beta}\Psi^{(2)}_{\alpha\rho} \tilde
T^{(2)\rho}_{(B^*2)\delta}\tilde
t^{(1)}_{(13)\beta}\epsilon^{\alpha\mu\delta\tau}\hat p_{\tau}
f^{B^*}_{B\gamma} \nonumber\\
 &&\hspace{1.4cm} + \;
 \frac{3}{5}C_{\bar{B}^*\bar{B}\gamma}\sum_{S_{\bar B^*}}i
\Psi^{C(2)}_{\alpha\rho}\Phi^{C(2)\nu\beta} \tilde
T^{(2)\rho}_{(\bar B^*1)\delta} \tilde
t^{(1)}_{(23)\beta}\epsilon^{\alpha\mu\delta\tau}\hat p_{\tau}
\bar f^{\bar B^*}_{\bar B \gamma}\;)\nonumber\\
 && \hspace{1.4cm}+ \;\sum_{S_{B^*}}
i\Phi^{(2)}_{\beta\lambda}\Psi^{(2)}_{\alpha\rho} \tilde
T^{(2)\rho}_{(B^*2)\delta} \tilde
t^{(3)\beta\lambda\nu}_{(13)}\epsilon^{\alpha\mu\delta\tau}\hat
p_{\tau}
f^{B^*}_{B\gamma}\nonumber\\
 &&\hspace{1.4cm} - \;\sum_{S_{\bar B^*}}i
\Psi^{C(2)}_{\alpha\rho} \Phi^{C(2)}_{\beta\lambda}\tilde
T^{(2)\rho}_{(\bar B^*1)\delta} \tilde
t^{(3)\beta\lambda\nu}_{(23)}\epsilon^{\alpha\mu\delta\tau}\hat
p_{\tau} \bar f^{\bar B^*}_{\bar B \gamma}\;)\;.
\end{eqnarray}
For $J/\psi (1^-) \to B^*({3\over 2}^-) \bar B({1\over 2}^-) +
\bar B^*({3\over 2}^+)  B({1\over 2}^+) \to \gamma B({1\over 2}^+)
\bar B({1\over 2}^-)$ we get the following six covariant
amplitudes
\begin{eqnarray}
(1,1,1)\nonumber\\
U^{\mu\nu}_{i,1} &=&
g^{i,a}_{J/\psi}\;g^{i,a}_{\gamma}(-\;\;\frac{2}{3}C_{B^*B\gamma}\sum_{S_{B^*}}
i\phi^{(1)\nu}\psi^{(1)}_\alpha \tilde
T^{(1)}_{(B^*2)\delta}\epsilon^{\alpha\mu\delta\tau}\hat
p_{\tau}f^{B^*}_{B\gamma}\nonumber\\
&&   \ \ \ \ \ \ \ \ \ \ \ \ +
\;\frac{2}{3}C_{\bar{B}^*\bar{B}\gamma}\sum_{S_{\bar B^*}}
i\psi^{C(1)}_\alpha \phi^{C(1)\nu} \tilde T^{(1)}_{(\bar B^*
1)\delta}\epsilon^{\alpha\mu\delta\tau}\hat p_{\tau}
 \bar f^{\bar B^*}_{\bar B \gamma}\nonumber\\
&&\ \ \ \ \ \ \ \ \ \ \ \   + \;\sum_{S_{B^*}}
i\phi^{(1)}_\beta\psi^{(1)}_\alpha \tilde T^{(1)}_{(B^*2)\delta}
\tilde t^{(2)\beta\nu}_{(13)}\epsilon^{\alpha\mu\delta\tau}\hat
p_{\tau}
f^{B^*}_{B\gamma}\nonumber\\
&&\ \ \ \ \ \ \ \ \ \ \ \  - \;\sum_{S_{\bar B^*}}i
\psi^{C(1)}_\alpha \phi^{C(1)}_\beta  \tilde T^{(1)}_{(\bar B^*
1)\delta} \tilde
t^{(2)\beta\nu}_{(23)}\epsilon^{\alpha\mu\delta\tau}\hat p_{\tau}
\bar f^{\bar B^*}_{\bar B \gamma}\;)\;, \\
(2,1,1)\nonumber\\
U^{\mu\nu}_{i,2} &=& g^{i,b}_{J/\psi}\;g^{i,a}_{\gamma} (-\;
\frac{2}{3}C_{B^*B\gamma}\sum_{S_{B^*}}\phi^{(1)\nu}
\Psi^{(2)\mu\alpha}\tilde T^{(1)}_{(B^*2)\alpha}
f^{B^*}_{B\gamma}\nonumber\\
&&\ \ \ \ \ \ \ \ \ \ \  -
\;\frac{2}{3}C_{\bar{B}^*\bar{B}\gamma}\sum_{S_{\bar B^*}}
\Psi^{C(2)\mu\alpha}\phi^{C(1)\nu}\tilde T^{(1)}_{(\bar B^*
1)\alpha} \bar f^{\bar B^*}_{\bar B \gamma}\nonumber\\
 && \ \ \ \ \ \ \ \ \ \ \ + \;\sum_{S_{B^*}}\phi^{(1)}_\beta
\Psi^{(2)\mu\alpha}\tilde T^{(1)}_{(B^*2)\alpha}
\tilde t^{(2)\beta\nu}_{(13)} f^{B^*}_{B\gamma}\nonumber\\
&&\ \ \ \ \ \ \ \ \ \  + \;\sum_{S_{\bar B^*}}
\Psi^{C(2)\mu\alpha}\tilde \phi^{C(1)}_\beta T^{(1)}_{(\bar B^*
1)\alpha}\tilde t^{(2)\beta\nu}_{(23)} \bar f^{\bar B^*}_{\bar B
\gamma}\;)
 \;,\\
(2,3,1)\nonumber\\
U^{\mu\nu}_{i,3} &=& g^{i,c}_{J/\psi}\;g^{i,a}_{\gamma}\;
(\;\sum_{S_{B^*}}-\frac{2}{3}C_{B^*B\gamma}\phi^{(1)\nu}\Psi^{(2)}_{\alpha\delta}
\tilde
T^{(3)\alpha\delta\mu}_{(B^*2)}f^{B^*}_{B\gamma}\nonumber\\
&&\ \ \ \ \ \ \ \ \ \ \ \ \ \ \ \ -
\;\frac{2}{3}C_{\bar{B}^*\bar{B}\gamma}\sum_{S_{\bar B^*}}
\Psi^{C(2)}_{\alpha\delta}\phi^{C(1)\nu} \tilde
T^{(3)\alpha\delta\mu}_{(\bar B^* 1)}\bar f^{\bar B^*}_{\bar B
\gamma}\nonumber\\
 && \ \ \ \ \ \ \ \ \ \ \ \ \ \ \ \ + \;
\sum_{S_{B^*}}\phi^{(1)}_\beta\Psi^{(2)}_{\alpha\delta} \tilde
T^{(3)\alpha\delta\mu}_{(B^*2)}
 \tilde t^{(2)\beta\nu}_{(13)}f^{B^*}_{B\gamma} \nonumber\\
&&\ \ \ \ \ \ \ \ \ \ \ \ \ \ \ \ +\;\sum_{S_{\bar B^*}}
 \Psi^{C(2)}_{\alpha\delta}\phi^{C(1)}_\beta \tilde
T^{(3)\alpha\delta\mu}_{(\bar B^* 1)}\tilde
t^{(2)\beta\nu}_{(23)}\bar f^{\bar B^*}_{\bar B \gamma}\;)
 \;,\\
(1,1,2)\nonumber\\
 U^{\mu\nu}_{i,4} &=&g^{i,a}_{J/\psi}\;g^{i,b}_{\gamma}\;
(-\sum_{S_{B^*}}\Phi^{(2)}_{\beta\xi}\psi^{(1)}_\alpha\epsilon^{\alpha\mu\delta\tau}
\tilde T^{(1)}_{(B^*2)\delta} \hat
p_{\tau}\epsilon^{\beta\nu\lambda\sigma} \tilde
t^{(2)\xi}_{(13)\lambda}\hat p_{B^*\sigma}f^{B^*}_{B\gamma} \nonumber\\
&&\ \ \ \ \ \ \ \ \ \ \  -\;\sum_{S_{\bar B^*}} \psi^{C(1)}_\alpha
\Phi^{C(2)}_{\beta\xi}\epsilon^{\alpha\mu\delta\tau} \tilde
T^{(1)}_{(\bar B^* 1)\delta} \hat
p_{\tau}\epsilon^{\beta\nu\lambda\sigma} \tilde
t^{(2)\xi}_{(23)\lambda}\hat p_{\bar B^*\sigma}\bar f^{\bar
B^*}_{\bar B \gamma}\;)
\;, \\
(2,1,2)\nonumber\\
U^{\mu\nu}_{i,5} &=&  \;g^{i,b}_{J/\psi}\;g^{i,b}_{\gamma}\;
(\;\sum_{S_{B^*}}i\Phi^{(2)}_{\beta\xi}\Psi^{(2)\mu\alpha}\tilde
T^{(1)}_{(B^*2)\alpha}\epsilon^{\beta\nu\lambda\sigma}
\tilde t^{(2)\xi}_{(13)\lambda}\hat p_{B^*\sigma}f^{B^*}_{B\gamma}\nonumber\\
&& \ \ \ \ \ \ \ \ \ \  -\;\sum_{S_{\bar
B^*}}i\Psi^{C(2)\mu\alpha}\Phi^{C(2)}_{\beta\xi}\tilde
T^{(1)}_{(\bar B^* 1)\alpha}\epsilon^{\beta\nu\lambda\sigma} \tilde
t^{(2)\xi}_{(23)\lambda}\hat p_{\bar B^*\sigma}\bar f^{\bar
B^*}_{\bar B \gamma}\;)
\;,\\
(2,3,2)\nonumber\\
U^{\mu\nu}_{i,6} &=&
\;g^{i,c}_{J/\psi}\;g^{i,b}_{\gamma}\;(\;\sum_{S_{B^*}}i\Phi^{(2)}_{\beta\xi}
\Psi^{(2)}_{\alpha\delta} \tilde
T^{(3)\alpha\delta\mu}_{(B^*2)}\epsilon^{\beta\nu\lambda\sigma}
\tilde t^{(2)\xi}_{(13)\lambda}\hat p_{B^*\sigma}f^{B^*}_{B\gamma}\nonumber\\
&& \ \ \ \ \ \ \ \ \ \ \ -\;\sum_{S_{\bar B^*}}i
\Psi^{C(2)}_{\alpha\delta}\Phi^{C(2)}_{\beta\xi} \tilde
T^{(3)\alpha\delta\mu}_{(\bar B^*
1)}\epsilon^{\beta\nu\lambda\sigma} \tilde
t^{(2)\xi}_{(23)\lambda}\hat p_{\bar B^*\sigma}\bar f^{\bar
B^*}_{\bar B \gamma}\;)
 \;.
\end{eqnarray}
For $J/\psi (1^-) \to B^*({5\over 2}^+) \bar B({1\over 2}^-) + \bar
B^*({5\over 2}^-) B({1\over 2}^+)\to \gamma B({1\over 2}^+) \bar
B({1\over 2}^-)$ we get the following six covariant amplitudes
\begin{eqnarray}
(2,2,2)\nonumber\\
U^{\mu\nu}_{i,1} &=& g^{i,a}_{J/\psi}\;g^{i,a}_{\gamma}\;
(-\;\frac{3}{5}C_{B^*B\gamma} \sum_{S_{B^*}}i\phi^{(2)\nu\beta}
\psi^{(2)}_{\alpha\rho}\epsilon^{\alpha\nu\delta\tau} \tilde
T^{(2)\rho}_{(B^*2)\delta }\hat p_\tau \tilde
t^{(1)}_{(13)\beta}f^{B^*}_{B\gamma}\nonumber\\ && \ \ \ \ \ \ \ \
\ \ \ \ \   +\;\frac{3}{5}C_{\bar{B}^*\bar{B}\gamma};\sum_{S_{\bar
B^*}}
i\psi^{C(2)}_{\alpha\rho}\phi^{C(2)\nu\beta}\epsilon^{\alpha\nu\delta\tau}
\tilde T^{(2)\rho}_{(\bar B^* 1)\delta} \hat p_\tau \tilde
t^{(1)}_{(23)\beta} \bar f^{\bar B^*}_{\bar B \gamma}\nonumber\\
&& \ \ \ \ \ \ \ \ \ \ \ \ \ + \; \sum_{S_{B^*}}i
\phi^{(2)}_{\beta\lambda}
\psi^{(2)}_{\alpha\rho}\epsilon^{\alpha\nu\delta\tau} \tilde
T^{(2)\rho}_{(B^*2)\delta }\hat p_\tau \tilde
t^{(3)\beta\lambda\nu}_{(13)}f^{B^*}_{B\gamma} \nonumber \\
&&   \ \ \ \ \ \ \ \ \ \ \ \ \ - \;\sum_{S_{\bar B^*}}
i\psi^{C(2)}_{\alpha\rho}\phi^{C(2)}_{\beta\lambda}\epsilon^{\alpha\nu\delta\tau}
\tilde T^{(2)\rho}_{(\bar B^* 1)\delta} \hat p_\tau \tilde
t^{(3)\beta\lambda\nu}_{(23)}\bar f^{\bar B^*}_{\bar B \gamma}\;)\;,
\nonumber\\ 
(3,2,2)\nonumber\\
U^{\mu\nu}_{i,2} &=& g^{i,b}_{J/\psi}\;g^{i,a}_{\gamma}
(-\frac{3}{5}C_{B^*B\gamma}
\sum_{S_{B^*}}\phi^{(2)\nu\beta}\Psi^{(3)\mu\alpha\delta}\tilde
T^{(2)}_{(B^*2)\alpha\delta}\tilde
t^{(1)}_{(13)\beta} f^{B^*}_{B\gamma} \nonumber\\
&&  \ \ \ \ \ \ \ \ \ - \;
\frac{3}{5}C_{\bar{B}^*\bar{B}\gamma}\;\sum_{S_{\bar B^*}}
\Psi^{C(3)\mu\alpha\delta} \phi^{C(2)\nu\beta}\tilde
T^{(2)}_{(\bar B^*1)\alpha\delta}\tilde t^{(1)}_{(23)\beta}\bar
f^{\bar B^*}_{\bar
B \gamma}\nonumber\\
&&  \ \ \ \ \ \ \ \ \  +\; \sum_{S_{B^*}}
\phi^{(2)}_{\beta\lambda}\Psi^{(3)\mu\alpha\delta}\tilde
T^{(2)}_{(B^*2)\alpha\delta}\tilde
t^{(3)\beta\lambda\nu}_{(13)}f^{B^*}_{B\gamma}
\nonumber\\
&&  \ \ \ \ \ \ \ \ \  + \;\sum_{S_{\bar
B^*}}\Psi^{C(3)\mu\alpha\delta}\phi^{C(2)}_{\beta\lambda}\tilde
T^{(2)}_{(\bar B^*1)\alpha\delta} \tilde
t^{(3)\beta\lambda\nu}_{(23)}\bar f^{\bar B^*}_{\bar B \gamma}\;)
\;,\\
(3,4,2)\nonumber\\
U^{\mu\nu}_{i,3} &=&
g^{i,c}_{J/\psi}\;g^{i,a}_{\gamma}(-\frac{3}{5}C_{B^*B\gamma}\sum_{S_{B^*}}\phi^{(2)\nu\beta}\Psi^{(3)}_{\alpha\delta\tau}\tilde
T^{(4)\alpha\delta\tau\mu}_{(B^*2)} \tilde
t^{(1)}_{(13)\beta}f^{B^*}_{B\gamma}\nonumber\\
&& \ \ \ \ \ \ \ \ \ \ \  -
\;\frac{3}{5}C_{\bar{B}^*\bar{B}\gamma} \sum_{S_{\bar
B^*}}\Psi^{C(3)}_{\alpha\delta\tau}\phi^{C(2)\nu\beta}\tilde
T^{(4)\alpha\delta\tau\mu}_{(\bar B^*1)}\tilde t^{(1)}_{(23)\beta}
\bar f^{\bar B^*}_{\bar B \gamma}\nonumber\\
&& \ \ \ \ \ \ \ \ \ \ \ + \;
\sum_{S_{B^*}}\phi^{(2)}_{\beta\lambda}\Psi^{(3)}_{\alpha\delta\tau}\tilde
T^{(4)\alpha\delta\tau\mu}_{(B^*2)} \tilde
t^{(3)\beta\lambda\nu}_{(13)}f^{B^*}_{B\gamma} \nonumber\\
&& \ \ \ \ \ \ \ \ \ \ \ + \;\sum_{S_{\bar B^*}}
\Psi^{C(3)}_{\alpha\delta\tau}\phi^{C(2)}_{\beta\lambda}\tilde
T^{(4)\alpha\delta\tau\mu}_{(\bar B^*1)} \tilde
t^{(3)\beta\lambda\nu}_{(23)}\bar f^{\bar B^*}_{\bar B \gamma}\;)
 \;,\\
(2,2,3)\nonumber\\
U^{\mu\nu}_{i,4} &=&g^{i,a}_{J/\psi}\;g^{i,b}_{\gamma}\;
(-\sum_{S_{B^*}}\Phi^{(3)}_{\beta\eta\xi}
\psi^{(2)}_{\alpha\rho}\epsilon^{\alpha\nu\delta\tau} \tilde
T^{(2)\rho}_{(B^*2)\delta} \hat p_\tau
\epsilon^{\beta\nu\lambda\sigma}
\tilde t^{(3)\eta\xi}_{(13)\lambda}\hat p_{B^*\sigma}f^{B^*}_{B\gamma}\nonumber\\
&& \ \ \ \ \ \ \ \ \ \ \ - \;\sum_{S_{\bar B^*}}
\psi^{C(2)}_{\alpha\rho}\Phi^{C(3)}_{\beta\eta\xi}\epsilon^{\alpha\nu\delta\tau}
\tilde T^{(2)\rho}_{(\bar B^* 1)\delta} \hat p_\tau
\epsilon^{\beta\nu\lambda\sigma} \tilde
t^{(3)\eta\xi}_{(23)\lambda}\hat p_{\bar B^*\sigma}\bar f^{\bar
B^*}_{\bar B \gamma}\;)
\;,\\
(3,2,3)\nonumber\\
 U^{\mu\nu}_{i,5} &=& g^{i,b}_{J/\psi}\;g^{i,b}_{\gamma}\;(\;\sum_{S_{B^*}}
i\Phi^{(3)}_{\beta\eta\xi}\Psi^{(3)\mu\alpha\delta}\tilde
T^{(2)}_{(B^*2)\alpha\delta} \epsilon^{\beta\nu\lambda\sigma} \tilde
t^{(3)\eta\xi}_{(13)\lambda}\hat p_{B^*\sigma}f^{B^*}_{B\gamma}\nonumber\\
&& \ \ \ \ \ \ \ \ \  - \; \sum_{S_{\bar B^*}}
i\Psi^{C(3)\mu\alpha\delta}\Phi^{C(3)}_{\beta\eta\xi}\tilde
T^{(2)}_{(\bar B^*1)\alpha\delta} \epsilon^{\beta\nu\lambda\sigma}
\tilde t^{(3)\eta\xi}_{(23)\lambda}\hat p_{\bar B^*\sigma}\bar
f^{\bar B^*}_{\bar B \gamma}\;)
\;,\\
(3,4,3)\nonumber\\
U^{\mu\nu}_{i,6} &=&  g^{i,c}_{J/\psi}\;g^{i,b}_{\gamma}\;
(\;\sum_{S_{B^*}}i\Phi^{(3)}_{\beta\eta\xi}\Psi^{(3)}_{\alpha\delta\tau}\tilde
T^{(4)\alpha\delta\tau\mu}_{(B^*2)} \epsilon^{\beta\nu\lambda\sigma}
\tilde
t^{(3)\eta\xi}_{(13)\lambda}\hat p_{B^*\sigma}f^{B^*}_{B\gamma} \nonumber\\
&& \ \ \ \ \ \ \ \ \   - \;\sum_{S_{\bar B^*}}i
\Psi^{C(3)}_{\alpha\delta\tau}\Phi^{C(3)}_{\beta\eta\xi}\tilde
T^{(4)\alpha\delta\tau\mu}_{(\bar B^* 1)}
\epsilon^{\beta\nu\lambda\sigma} \tilde
t^{(3)\eta\xi}_{(23)\lambda}\hat p_{\bar B^*\sigma} \bar f^{\bar
B^*}_{\bar B \gamma}\;)\;.
\end{eqnarray}
For $J/\psi (1^-) \to B^*({5\over 2}^-) \bar B({1\over 2}^-) + \bar
B^*({5\over 2}^+) \bar B({1\over 2}^+) \to \gamma B({1\over 2}^+)
\bar B({1\over 2}^-)$ we get the following six covariant amplitudes
\begin{eqnarray}
(2,1,,2)\nonumber\\
U^{\mu\nu}_{i,1} &=& g^{i,a}_{J/\psi}\;g^{i,a}_{\gamma}\;
(\;\sum_{S_{B^*}}i\phi^{(2)}_{\beta\eta}\psi^{(2)\mu\alpha}\tilde
T^{(1)}_{(B^*2)\alpha} \epsilon^{\beta\nu\lambda\sigma}
\tilde t^{(2)\eta}_{(13)\lambda}\hat p_{B^*\sigma}f^{B^*}_{B\gamma} \nonumber\\
&& \ \ \ \ \ \ \ \ \ \ \  - \; \sum_{S_{\bar B^*}}
i\psi^{C(2)\mu\alpha}\phi^{C(2)}_{\beta\eta}\tilde T^{(1)}_{(\bar
B^*1)\alpha} \epsilon^{\beta\nu\lambda\sigma} \tilde
t^{(2)\eta}_{(23)\lambda}\hat p_{\bar B^*\sigma}\bar f^{\bar
B^*}_{\bar B \gamma}\;)
\;,\\
(2,3,2)\nonumber\\
U^{\mu\nu}_{i,2} &=& g^{i,b}_{J/\psi}\;g^{i,a}_{\gamma}\;
(\;\sum_{S_{B^*}}i\phi^{(2)}_{\beta\eta}\psi^{(2)}_{\alpha\delta}
\tilde T^{(3)\alpha\delta\mu}_{(B^*2)}
\epsilon^{\beta\nu\lambda\sigma}
\tilde t^{(2)\eta}_{(13)\lambda}\hat p_{B^*\sigma} f^{B^*}_{B\gamma}\nonumber\\
&& \ \ \ \ \ \ \ \ \ \ \  - \;\sum_{S_{\bar B^*}}i
\psi^{C(2)}_{\alpha\delta}\phi^{C(2)}_{\beta\eta} \tilde
T^{(3)\alpha\delta\mu}_{(\bar B^*1)}
\epsilon^{\beta\nu\lambda\sigma} \tilde
t^{(2)\eta}_{(23)\lambda}\hat p_{\bar B^*\sigma}\bar f^{\bar
B^*}_{\bar B \gamma}\;)
\;,\\
(3,3,2)\nonumber\\
U^{\mu\nu}_{i,3} &=&
g^{i,c}_{J/\psi}\;g^{i,a}_{\gamma}\;(-\sum_{S_{B^*}}
\phi^{(2)}_{\beta\eta}\Psi^{(3)}_{\alpha\rho\zeta}\epsilon^{\alpha\mu\delta\tau}
\tilde T^{(3)\rho\zeta}_{(B^*2)\delta}\hat p_\tau
\epsilon^{\beta\nu\lambda\sigma}
\tilde t^{(2)\eta}_{(13)\lambda}\hat p_{B^*\sigma}f^{B^*}_{B\gamma} \nonumber\\
&& \ \ \ \ \ \ \  -\;\sum_{S_{\bar B^*}}
\Psi^{C(3)}_{\alpha\rho\zeta}\phi^{C(2)}_{\beta\eta}\epsilon^{\alpha\mu\delta\tau}
\tilde T^{(3)\rho\zeta}_{(\bar B^*1)\delta}\hat p_\tau
\epsilon^{\beta\nu\lambda\sigma} \tilde
t^{(2)\eta}_{(23)\lambda}\hat p_{\bar B^*\sigma}\bar f^{\bar
B^*}_{\bar B \gamma}\;)
\;,\\
(2,1,3)\nonumber\\
U^{\mu\nu}_{i,4} &=& g^{i,a}_{J/\psi}\;g^{i,b}_{\gamma}\;
(-\frac{4}{7}C_{B^*B\gamma}
\sum_{S_{B^*}}\Phi^{(3)\nu\lambda\sigma}\psi^{(2)\mu\alpha}\tilde
T^{(1)}_{(B^*2)\alpha} \tilde
t^{(2)}_{(13)\lambda\sigma}f^{B^*}_{B\gamma} \nonumber\\
&& \ \ \ \ \ \ \ \ \ \ \ \ -
\;\frac{4}{7}C_{\bar{B}^*\bar{B}\gamma}\sum_{S_{\bar B^*}}
\psi^{C(2)\mu\alpha}\Phi^{C(3)\nu\lambda\sigma}\tilde
T^{(1)}_{(\bar B^*1)\alpha} \tilde t^{(2)}_{(23)\lambda\sigma}
\bar f^{\bar
B^*}_{\bar B \gamma}\nonumber\\
&& \ \ \ \ \ \ \ \ \ \ \ \  + \; \sum_{S_{B^*}}
\Phi^{(3)}_{\beta\lambda\sigma}\psi^{(2)\mu\alpha}\tilde
T^{(1)}_{(B^*2)\alpha} \tilde
t^{(4)\beta\lambda\sigma\nu}_{(13)}f^{B^*}_{B\gamma}\nonumber\\
&& \ \ \ \ \ \ \ \ \ \ \ \  + \; \sum_{S_{\bar B^*}}
\psi^{C(2)\mu\alpha}\Phi^{C(3)}_{\beta\lambda\sigma}\tilde
T^{(1)}_{(\bar B^*1)\alpha} \tilde
t^{(4)\beta\lambda\sigma\nu}_{(23)} \bar f^{\bar B^*}_{\bar B
\gamma}\;),\\
(2,3,3)\nonumber\\
U^{\mu\nu}_{i,5} &=& g^{i,b}_{J/\psi}\;g^{i,b}_{\gamma}\;
(-\frac{4}{7}C_{B^*B\gamma}\sum_{S_{B^*}}\Phi^{(3)\nu\lambda\sigma}\psi^{(2)}_{\alpha\delta}
\tilde T^{(3)\alpha\delta\mu}_{(B^*2)}\tilde
t^{(2)}_{(13)\lambda\sigma}f^{B^*}_{B\gamma} \nonumber\\
&& \ \ \ \ \ \ \ \ \ \ \  -\frac{4}{7}C_{\bar{B}^*\bar{B}\gamma}
\;\sum_{S_{\bar B^*}}
\psi^{C(2)}_{\alpha\delta}\Phi^{C(3)\nu\lambda\sigma} \tilde
T^{(3)\alpha\delta\mu}_{(\bar B^*1)}\tilde
t^{(2)}_{(23)\lambda\sigma} \bar f^{\bar B^*}_{\bar B \gamma}
\nonumber\\
&& \ \ \ \ \ \ \ \ \ \ \ + \;
\sum_{S_{B^*}}\Phi^{(3)}_{\beta\lambda\sigma}\psi^{(2)}_{\alpha\delta}
\tilde T^{(3)\alpha\delta\mu}_{(B^*2)} \tilde
t^{(4)\beta\lambda\sigma\nu}_{(13)} f^{B^*}_{B\gamma}\nonumber\\
 && \ \ \ \ \ \ \ \ \ \ \   + \;\sum_{S_{\bar B^*}}
\psi^{C(2)}_{\alpha\delta}\Phi^{C(3)}_{\beta\lambda\sigma} \tilde
T^{(3)\alpha\delta\mu}_{(\bar B^*1)}\tilde
t^{(4)\beta\lambda\sigma\nu}_{(23)}\bar f^{\bar B^*}_{\bar B \gamma}
\;)
\;,\\
(3,3,3)\nonumber\\
U^{\mu\nu}_{i,6} &=& g^{i,c}_{J/\psi}\;g^{i,b}_{\gamma}\;
(-\;\frac{4}{7}C_{B^*B\gamma}\sum_{S_{B^*}}i\Phi^{(3)\nu\lambda\sigma}
\Psi^{(3)}_{\alpha\rho\zeta}\epsilon^{\alpha\mu\delta\tau} \tilde
T^{(3)\rho\zeta}_{(B^*2)\delta}\hat p_\tau
\tilde t^{(2)}_{(13)\lambda\sigma}f^{B^*}_{B\gamma} \nonumber\\
\nonumber\\
 && \ \ \ \ \ \ \ \ \ \ \   -\;\;\frac{4}{7}C_{\bar{B}^*\bar{B}\gamma}\; \sum_{S_{\bar B^*}}
 i\Psi^{C(3)}_{\alpha\rho\zeta}\Phi^{C(3)\nu\lambda\sigma}
 \epsilon^{\alpha\mu\delta\tau}
\tilde T^{(3)\rho\zeta}_{(\bar B^*1)\delta}\hat p_\tau
\tilde t^{(2)}_{(23)\lambda\sigma}\bar f^{\bar B^*}_{\bar B \gamma}
\nonumber\\
&&  \ \ \ \ \ \ \ \ \ \ \ \ \  +
\sum_{S_{B^*}}i\Phi^{(3)}_{\beta\lambda\sigma}
\Psi^{(3)}_{\alpha\rho\zeta}\epsilon^{\alpha\mu\delta\tau} \tilde
T^{(3)\rho\zeta}_{(B^*2)\delta}\hat p_\tau \tilde
t^{(4)\beta\lambda\sigma\nu}_{(13)}f^{B^*}_{B\gamma} \nonumber\\
\nonumber\\
 && \ \ \ \ \ \ \ \ \ \ \ \ \ + \;\sum_{S_{\bar B^*}}i
\Psi^{C(3)}_{\alpha\rho\zeta}\Phi^{C(3)}_{\beta\lambda\sigma}
\epsilon^{\alpha\mu\delta\tau} \tilde T^{(3)\rho\zeta}_{(\bar
B^*1)\delta}\hat p_\tau \tilde t^{(4)\beta\lambda\sigma\nu}_{(23)}\;
\bar f^{\bar B^*}_{\bar B \gamma}) \;.
\end{eqnarray}
For $J/\psi (1^-) \to B^*({7\over 2}^+) \bar B({1\over 2}^-) + \bar
B^*({7\over 2}^-) B({1\over 2}^+) \to \gamma B({1\over 2}^+) \bar
B({1\over 2}^-)$ we get the following six covariant amplitudes
\begin{eqnarray}
(3,2,3)\nonumber\\
U^{\mu\nu}_{i,1} &=&
g^{i,a}_{J/\psi}\;g^{i,a}_{\gamma}\;(\;\sum_{S_{B^*}}i
\phi^{(3)}_{\beta\eta\xi}\psi^{(3)\mu\alpha\delta}\tilde
T^{(2)}_{(B^*2)\alpha\delta} \epsilon^{\beta\nu\lambda\sigma}
\tilde t^{(3)\eta\xi}_{(13)\lambda}\hat p_{B^*\sigma}f^{B^*}_{B\gamma}\nonumber\\
&& \ \ \ \ \ \ \ \ \  -\; \sum_{S_{\bar
B^*}}i\psi^{C(3)\mu\alpha\delta}\phi^{C(3)}_{\beta\eta\xi}\tilde
T^{(2)}_{(\bar B^*1)\alpha\delta} \epsilon^{\beta\nu\lambda\sigma}
\tilde t^{(3)\eta\xi}_{(23)\lambda}\hat p_{\bar B^*\sigma}\bar
f^{\bar B^*}_{\bar B \gamma}\;)
 \;,\\
(3,4,3)\nonumber\\
U^{\mu\nu}_{i,2} &=&
g^{i,b}_{J/\psi}\;g^{i,a}_{\gamma}\;(\;\sum_{S_{B^*}}
i\phi^{(3)}_{\beta\eta\xi}\psi^{(3)}_{\alpha\delta\tau}\tilde
T^{(4)\alpha\delta\tau\mu}_{(B^*2)} \epsilon^{\beta\nu\lambda\sigma}
\tilde t^{(3)\eta\xi}_{(13)\lambda}\hat p_{B^*\sigma}f^{B^*}_{B\gamma} \nonumber\\
&& \ \ \ \ \ \ \ \ \  -\; \sum_{S_{\bar B^*}}i
\psi^{C(3)}_{\alpha\delta\tau}\phi^{C(3)}_{\beta\eta\xi}\tilde
T^{(4)\alpha\delta\tau\mu}_{(\bar B^*1)}
\epsilon^{\beta\nu\lambda\sigma} \tilde
t^{(3)\eta\xi}_{(23)\lambda}\hat p_{\bar B^*\sigma}\bar f^{\bar
B^*}_{\bar B \gamma}\;)
 \;,\\
(4,4,3)\nonumber\\
U^{\mu\nu}_{i,3} &=&
g^{i,c}_{J/\psi}\;g^{i,a}_{\gamma}\;(-\sum_{S_{B^*}}
\phi^{(3)}_{\beta\eta\xi}\Psi^{(4)}_{\alpha\rho\zeta\gamma}
\epsilon^{\alpha\mu\delta\tau} \tilde
T^{(4)\rho\zeta\gamma}_{(B^*2)\delta} \hat p_\tau
\epsilon^{\beta\nu\lambda\sigma}
\tilde t^{(3)\eta\xi}_{(13)\lambda}\hat p_{B^*\sigma}f^{B^*}_{B\gamma} \nonumber\\
&& \ \ \ \ \ \ \ \ \ \ \ - \;\sum_{S_{\bar B^*}}
\Psi^{C(4)}_{\alpha\rho\zeta\gamma}\phi^{C(3)}_{\beta\eta\xi}
\epsilon^{\alpha\mu\delta\tau} \tilde T^{(4)\rho\zeta\gamma}_{(\bar
B^*1)\delta} \hat p_\tau \epsilon^{\beta\nu\lambda\sigma} \tilde
t^{(3)\eta\xi}_{(23)\lambda}\hat p_{\bar B^*\sigma}\bar f^{\bar
B^*}_{\bar B \gamma}\;)
 \;,\\
(3,2,4)\nonumber\\
U^{\mu\nu}_{i,4} &=&
g^{i,a}_{J/\psi}\;g^{i,b}_{\gamma}(-\frac{5}{9}C_{B^*B\gamma}
\sum_{S_{B^*}}
\Phi^{(4)\nu\lambda\sigma\eta}\psi^{(3)\mu\alpha\delta}\tilde
T^{(2)}_{(B^*2)\alpha\delta} \tilde
t^{(3)}_{(13)\lambda\sigma\eta}f^{B^*}_{B\gamma} \nonumber\\
&& \ \ \ \ \ \ \ \ \   -
\;\frac{5}{9}C_{\bar{B}^*\bar{B}\gamma}\sum_{S_{\bar B^*}}
\psi^{C(3)\mu\alpha\delta}\Phi^{C(4)\nu\lambda\sigma\eta}\tilde
T^{(2)}_{(\bar B^*1)\alpha\delta} \tilde
t^{(3)}_{(23)\lambda\sigma\eta} \bar f^{\bar B^*}_{\bar B \gamma}\nonumber\\
&&\ \ \ \ \ \ \ \ \ + \;
\;\sum_{S_{B^*}}\Phi^{(4)}_{\beta\lambda\sigma\eta}\psi^{(3)\mu\alpha\delta}\tilde
T^{(2)}_{(B^*2)\alpha\delta} \tilde
t^{(5)\beta\lambda\sigma\eta\nu}_{(13)} f^{B^*}_{B\gamma} \nonumber\\
&& \ \ \ \ \ \ \ \ \  + \; \sum_{S_{\bar
B^*}}\psi^{C(3)\mu\alpha\delta}\Phi^{C(4)}_{\beta\lambda\sigma\eta}\tilde
T^{(2)}_{(\bar B^*1)\alpha\delta} \tilde
t^{(5)\beta\lambda\sigma\eta\nu}_{(23)}\bar f^{\bar B^*}_{\bar B
\gamma}\;)
 \;,\\
(3,4,4)\nonumber\\
U^{\mu\nu}_{i,5} &=&
g^{i,b}_{J/\psi}\;g^{i,b}_{\gamma}\;(-\;\frac{5}{9}C_{B^*B\gamma}\sum_{S_{B^*}}
\Phi^{(4)\nu\lambda\sigma\eta}\psi^{(3)}_{\alpha\delta\tau}\tilde
T^{(4)\alpha\delta\tau\mu}_{(B^*2)} \tilde
t^{(3)}_{(13)\lambda\sigma\eta} f^{B^*}_{B\gamma} \nonumber\\
&& \ \ \ \ \ \ \ \ \ \ \ \  - \;
\frac{5}{9}C_{\bar{B}^*\bar{B}\gamma}\;\sum_{S_{\bar B^*}}
\psi^{C(3)}_{\alpha\delta\tau}\Phi^{C(4)\nu\lambda\sigma\eta}\tilde
T^{(4)\alpha\delta\tau\mu}_{(\bar B^*1)} \tilde
t^{(3)}_{(23)\lambda\sigma\eta} \bar f^{\bar B^*}_{\bar B \gamma}\nonumber\\
&& \ \ \ \ \ \ \ \ \ \ \ \ + \;
\sum_{S_{B^*}}\Phi^{(4)}_{\beta\lambda\sigma\eta}
\psi^{(3)}_{\alpha\delta\tau}\tilde
T^{(4)\alpha\delta\tau\mu}_{(B^*2)} \tilde
t^{(5)\beta\lambda\sigma\eta\nu}_{(13)} f^{B^*}_{B\gamma} \nonumber\\
&& \ \ \ \ \ \ \ \ \ \ \ \ + \; \sum_{S_{\bar
B^*}}\psi^{C(3)}_{\alpha\delta\tau}\Phi^{C(4)}_{\beta\lambda\sigma\eta}\tilde
T^{(4)\alpha\delta\tau\mu}_{(\bar B^*1)} \tilde
t^{(5)\beta\lambda\sigma\eta\nu}_{(23)}\bar f^{\bar B^*}_{\bar B
\gamma}\;)
 \;,\\
(4,4,4)\nonumber\\
U^{\mu\nu}_{i,6} &=& g^{i,c}_{J/\psi}\;g^{i,b}_{\gamma}\; (- \;
\frac{5}{9}C_{B^*B\gamma}\sum_{S_{B^*}}i
\Phi^{(4)\nu\lambda\sigma\eta}
\Psi^{(4)}_{\alpha\rho\zeta\gamma}\epsilon^{\alpha\mu\delta\tau}
\tilde T^{(4)\rho\zeta\gamma}_{(B^*2)\delta}\hat p_\tau \tilde
t^{(3)}_{(13)\lambda\sigma\eta}f^{B^*}_{B\gamma}
\nonumber\\
&& \ \ \ \ \ \ \ \ \ \ \ \ \ + \;
\frac{5}{9}C_{\bar{B}^*\bar{B}\gamma}\sum_{S_{\bar B^*}}i
\Psi^{C(4)}_{\alpha\rho\zeta\gamma}\Phi^{C(4)\nu\lambda\sigma\eta}
\epsilon^{\alpha\mu\delta\tau} \tilde
T^{(4)\rho\zeta\gamma}_{(\bar B^*2)\delta}\hat p_\tau \tilde
t^{(3)}_{(23)\lambda\sigma\eta}\bar
f^{\bar B^*}_{\bar B \gamma}\nonumber\\
 && \ \ \ \ \ \ \ \ \ \ \ \ \ + \; \sum_{S_{B^*}}i\Phi^{(4)}_{\beta\lambda\sigma\eta}
i\Psi^{(4)}_{\alpha\rho\zeta\gamma}\epsilon^{\alpha\mu\delta\tau}
\tilde T^{(4)\rho\zeta\gamma}_{(B^*2)\delta}\hat p_\tau \tilde
t^{(5)\beta\lambda\sigma\eta\nu}_{(13)} f^{B^*}_{B\gamma}\nonumber\\
&& \ \ \ \ \ \ \ \ \ \ \ \ \ - \;\sum_{S_{\bar B^*}}i
\Psi^{C(4)}_{\alpha\rho\zeta\gamma}\Phi^{C(4)}_{\beta\lambda\sigma\eta}
\epsilon^{\alpha\mu\delta\tau} \tilde T^{(4)\rho\zeta\gamma}_{(\bar
B^*2)\delta}\hat p_\tau \tilde
t^{(5)\beta\lambda\sigma\eta\nu}_{(23)}\bar f^{\bar B^*}_{\bar B
\gamma}\;)
 \;.
\end{eqnarray}
For $J/\psi (1^-) \to B^*({7\over 2}^-) \bar B({1\over 2}^-) + \bar
B^*({7\over 2}^+)  B({1\over 2}^+) \to \gamma B({1\over 2}^+) \bar
B({1\over 2}^-)$ we get the following six covariant amplitudes
\begin{eqnarray}
(3,3,3)\nonumber\\
U^{\mu\nu}_{i,1} &=& g^{i,a}_{J/\psi}\;g^{i,a}_{\gamma}\;
(-\frac{4}{7}C_{B^*B\gamma}\sum_{S_{B^*}}i\phi^{(3)\nu\beta\lambda}
\psi^{(3)}_{\alpha\rho\zeta}\epsilon^{\alpha\mu\delta\tau} \tilde
T^{(3)\rho\zeta}_{(B^*2)\delta} \hat p_\tau
\tilde t^{(2)}_{(13)\beta\lambda} f^{B^*}_{B\gamma} \nonumber\\
&& \ \ \ \ \ \ \ \   +\;
\frac{4}{7}C_{\bar{B}^*\bar{B}\gamma}\;\sum_{S_{\bar B^*}}i
\psi^{C(3)}_{\alpha\rho\zeta}\phi^{C(3)\nu\beta\lambda}
\epsilon^{\alpha\mu\delta\tau} \tilde T^{(3)\rho\zeta}_{(\bar
B^*1)\delta} \hat p_\tau \tilde t^{(2)}_{(23)\beta\lambda}\bar
f^{\bar B^*}_{\bar B \gamma} \nonumber\\
&&  \ \ \ \ \ \ \ \  + \;
\sum_{S_{B^*}}\phi^{(3)}_{\beta\lambda\sigma}i
\psi^{(3)}_{\alpha\rho\zeta}\epsilon^{\alpha\mu\delta\tau} \tilde
T^{(3)\rho\zeta}_{(B^*2)\delta} \hat p_\tau \tilde
t^{(4)\beta\lambda\sigma\nu}_{(13)}f^{B^*}_{B\gamma} \nonumber\\
&&  \ \ \ \ \ \ \ \   - \;\sum_{S_{\bar B^*}}i
\psi^{C(3)}_{\alpha\rho\zeta}
\phi^{C(3)}_{\beta\lambda\sigma}\epsilon^{\alpha\mu\delta\tau}
\tilde T^{(3)\rho\zeta}_{(\bar B^*1)\delta} \hat p_\tau \tilde
t^{(4)\beta\lambda\sigma\nu}_{(23)}\bar f^{\bar B^*}_{\bar B
\gamma}\;)
\;, \\
(4,3,3)\nonumber\\
U^{\mu\nu}_{i,2} &=&
g^{i,b}_{J/\psi}\;g^{i,a}_{\gamma}\;(-\frac{4}{7}C_{B^*B\gamma}\sum_{S_{B^*}}
\phi^{(3)\nu\beta\lambda}\Psi^{(4)\mu\alpha\delta\tau} \tilde
T^{(3)}_{(B^*2)\alpha\delta\tau} \tilde
t^{(2)}_{(13)\beta\lambda} f^{B^*}_{B\gamma} \nonumber\\
&& \ \ \ \ \ \ \ \ \ \ \
-\frac{4}{7}C_{\bar{B}^*\bar{B}\gamma}\;\sum_{S_{\bar B^*}}
\Psi^{C(4)\mu\alpha\delta\tau}\phi^{C(3)\nu\beta\lambda} \tilde
T^{(3)}_{(\bar B^*1)\alpha\delta\tau} \tilde
t^{(2)}_{(23)\beta\lambda} \bar f^{\bar B^*}_{\bar B
\gamma}\nonumber\\&& \ \ \ \ \ \ \ \ \ \ \
+\;\sum_{S_{B^*}}\phi^{(3)}_{\beta\lambda\sigma}\Psi^{(4)\mu\alpha\delta\tau}
\tilde T^{(3)}_{(B^*2)\alpha\delta\tau} \tilde
t^{(4)\beta\lambda\sigma\nu}_{(13)}f^{B^*}_{B\gamma} \nonumber\\
&& \ \ \ \ \ \ \ \ \ \ \  + \;\sum_{S_{\bar B^*}}
\Psi^{C(4)\mu\alpha\delta\tau}\phi^{C(3)}_{\beta\lambda\sigma}
\tilde T^{(3)}_{(\bar B^*1)\alpha\delta\tau} \tilde
t^{(4)\beta\lambda\sigma\nu}_{(23)}\bar f^{\bar B^*}_{\bar B
\gamma}\;)
\;, \\
(4,5,3)\nonumber\\
U^{\mu\nu}_{i,3} &=&
g^{i,c}_{J/\psi}\;g^{i,a}_{\gamma}\;(-\frac{4}{7}C_{B^*B\gamma}\sum_{S_{B^*}}
\phi^{(3)\nu\beta\lambda}\Psi^{(4)}_{\alpha\delta\tau\rho}\tilde
T^{(5)\alpha\delta\tau\rho\mu}_{(B^*2)}
\tilde t^{(2)}_{(13)\beta\lambda}f^{B^*}_{B\gamma} \nonumber\\
&& \ \ \ \ \ \ \ \ \ \ \
-\frac{4}{7}C_{\bar{B}^*\bar{B}\gamma}\;\sum_{S_{\bar B^*}}
\Psi^{C(4)}_{\alpha\delta\tau\rho}\phi^{C(3)\nu\beta\lambda}\tilde
T^{(5)\alpha\delta\tau\rho\mu}_{(\bar B^*1)} \tilde
t^{(2)}_{(23)\beta\lambda}\bar f^{\bar B^*}_{\bar B \gamma}\nonumber\\
&&\ \ \ \ \ \ \ \ \ \ \ +
\;\sum_{S_{B^*}}\phi^{(3)}_{\beta\lambda\sigma}
\Psi^{(4)}_{\alpha\delta\tau\rho}\tilde
T^{(5)\alpha\delta\tau\rho\mu}_{(B^*2)} \tilde
t^{(4)\beta\lambda\sigma\nu}_{(13)}f^{B^*}_{B\gamma}\nonumber\\
&& \ \ \ \ \ \ \ \ \ \ \ + \;\sum_{S_{\bar
B^*}}\Psi^{C(4)}_{\alpha\delta\tau\rho}\phi^{C(3)}_{\beta\lambda\sigma}\tilde
T^{(5)\alpha\delta\tau\rho\mu}_{(\bar B^*1)} \tilde
t^{(4)\beta\lambda\sigma\nu}_{(23)}\bar f^{\bar B^*}_{\bar B
\gamma}\;)
\;, \\
(3,3,4)\nonumber\\
U^{\mu\nu}_{i,4} &=&
g^{i,a}_{J/\psi}\;g^{i,b}_{\gamma}\;(-\sum_{S_{B^*}}
\Phi^{(4)}_{\beta\eta\xi\varrho}
\psi^{(3)}_{\alpha\rho\zeta}\epsilon^{\alpha\mu\delta\tau} \tilde
T^{(3)\rho\zeta}_{(B^*2)\delta}\hat p_\tau
\epsilon^{\beta\nu\lambda\sigma}
\tilde t^{(4)\eta\xi\varrho}_{(13)\lambda} \hat p_{B^*\sigma}f^{B^*}_{B\gamma}\nonumber\\
&& \ \ \ \ \ \ \ \  - \;\sum_{S_{\bar B^*}}
\psi^{C(3)}_{\alpha\rho\zeta}\Phi^{C(4)}_{\beta\eta\xi\varrho}
\epsilon^{\alpha\mu\delta\tau} \tilde T^{(3)\rho\zeta}_{(\bar
B^*1)\delta}\hat p_\tau \epsilon^{\beta\nu\lambda\sigma} \tilde
t^{(4)\eta\xi\varrho}_{(23)\lambda} \hat p_{\bar B^*\sigma}\bar
f^{\bar B^*}_{\bar B \gamma}\;)
 \;, \\
(4,4,4)\nonumber\\
U^{\mu\nu}_{i,5} &=&
g^{i,b}_{J/\psi}\;g^{i,b}_{\gamma}\;(\;\sum_{S_{B^*}}
i\Phi^{(4)}_{\beta\eta\xi\varrho}\Psi^{(4)\mu\alpha\delta\tau}
\tilde T^{(3)}_{(B^*2)\alpha\delta\tau}
\epsilon^{\beta\nu\lambda\sigma}
\tilde t^{(4)\eta\xi\varrho}_{(13)\lambda}\hat p_{B^*\sigma}f^{B^*}_{B\gamma}\nonumber\\
&& \ \ \ \ \ \ \ \ \  - \; \sum_{S_{\bar B^*}}
i\Psi^{C(4)\mu\alpha\delta\tau}\Phi^{C(4)}_{\beta\eta\xi\varrho}
\tilde T^{(3)}_{(\bar B^*2)\alpha\delta\tau}
\epsilon^{\beta\nu\lambda\sigma}
\tilde t^{(4)\eta\xi\varrho}_{(23)\lambda}\hat p_{\bar B^*\sigma}\bar f^{\bar B^*}_{\bar B \gamma} \;)\;, \\
(4,5,4)\nonumber\\
U^{\mu\nu}_{i,6} &=&
g^{i,c}_{J/\psi}\;g^{i,b}_{\gamma}\;(\;\sum_{S_{B^*}}i
\Phi^{(4)}_{\beta\eta\xi\varrho}\Psi^{(4)}_{\alpha\delta\tau\rho}\tilde
T^{(5)\alpha\delta\tau\rho\mu}_{(B^*2)}
\epsilon^{\beta\nu\lambda\sigma}
\tilde t^{(4)\eta\xi\varrho}_{(13)\lambda}\hat p_{B^*\sigma}f^{B^*}_{B\gamma}\nonumber\\
&& \ \ \ \ \ \ \ \ \ \   - \;\sum_{S_{\bar B^*}}i
\Psi^{C(4)}_{\alpha\delta\tau\rho}\Phi^{C(4)}_{\beta\eta\xi\varrho}\tilde
T^{(5)\alpha\delta\tau\rho\mu}_{(\bar B^*1)}
\epsilon^{\beta\nu\lambda\sigma} \tilde
t^{(4)\eta\xi\varrho}_{(23)\lambda}\hat p_{\bar B^*\sigma}\bar
f^{\bar B^*}_{\bar B \gamma}\;)
 \;.
\end{eqnarray}
Note that where
\begin{eqnarray*}
\psi^{(n)}_{\mu_1\cdots\mu_n} &=&
\psi^{(n)}_{\mu_1\cdots\mu_n}(p_{B^*},S_{B^*};p_{\bar B},S_{\bar
B})\;,\hspace{0.3cm} \Psi^{(n+1)}_{\mu_1\cdots\mu_{n+1}} =
\Psi^{(n+1)}_{\mu_1\cdots\mu_{n+1}} (p_{B^*},S_{B^*};p_{\bar
B},S_{\bar B})\;,\\
 \phi^{(n)}_{\mu_1\cdots\mu_n} &=&
\phi^{(n)}_{\mu_1\cdots\mu_{n+}}(p_{B^*},S_{B^*};p_B,S_B)\;,\hspace{0.3cm}
\Phi^{(n+1)}_{\mu_1\cdots\mu_{n+}} =
\Phi^{(n+1)}_{\mu_1\cdots\mu_{n+}}(p_{B^*},S_{B^*};p_B,S_B)\;,\\
\psi^{C(n)}_{\mu_1\cdots\mu_n} &=&
\psi^{C(n)}_{\mu_1\cdots\mu_n}(p_{\bar B^*},S_{\bar
B^*};p_B,S_B)\;,\hspace{0.3cm} \Psi^{C(n+1)}_{\mu_1\cdots\mu_{n+1}}
=\Psi^{C(n+1)}_{\mu_1\cdots\mu_{n+1}}(p_{\bar B^*},S_{\bar
B^*};p_B,S_B)\;,\\
 \phi^{C(n)}_{\mu_1\cdots\mu_n} &=& \phi^{C(n)}_{\mu_1\cdots\mu_n}(p_{\bar
B^*},S_{\bar B^*};p_{\bar B},S_{\bar B})\;,\hspace{0.3cm}
 \Phi^{C(n+1)}_{\mu_1\cdots\mu_{n+1}} = \Phi^{C(n+1)}_{\mu_1\cdots\mu_{n+1}}(p_{\bar
B^*},S_{\bar B^*};p_{\bar B},S_{\bar B})\;.
\end{eqnarray*}
and
\begin{eqnarray}
 C_{B^*B\gamma}&=&-\frac{(m^2_{B^*}-m^2_{B})^2}{m^2_{B^*}}\;,\hspace{0.3cm}
 C_{\bar{B}^*\bar{B}\gamma}=-\frac{(m^2_{\bar{B}^*}-m^2_{\bar{B}})^2}{m^2_{\bar{B}^*}}\;\nonumber.
\end{eqnarray}

For the reaction $J/\psi\to p \bar p \to \gamma p\bar p$ we get
following two covariant amplitudes with two independent coupling
constants $g^{p,a}$ and $g^{p,b}$ which are determined by the
experiment.
\begin{eqnarray}
(1,0,1)\nonumber\\
 U^{\mu\nu}_{p,1} &=&  g^{p,a}\;
 (\; \sum_{S_{p^*}}\Phi^{\nu}_{p^*} \Psi^{(1)\mu}f^{p^*}_{p\gamma} - \;\sum_{S_{\bar p^*}}\Psi^{C(1)\mu}\Phi^{C\nu}_{\bar p^*}f^{\bar p^*}_{\bar p \gamma}),\\
(1,2,1)\nonumber\\
 U^{\mu\nu}_{p,2} &=&  g^{p,b}\; (\;\sum_{S_{p^*}} \Phi^{\nu}_{p^*}\Psi^{(1)}_{\alpha}\; \tilde T^{(2)\alpha\mu}_{(p^*2)}f^{p^*}_{p\gamma}
 - \;\sum_{S_{\bar p^*}} \Psi^{C(1)}_{\alpha}\Phi^{C\nu}_{\bar p^*}
  \tilde T^{(2)\alpha\mu}_{(\bar p^*1)}\bar f^{\bar p^*}_{\bar p \gamma}\; ) \;,
\end{eqnarray}
 here we use $p^*$ and $\bar p^*$ as intermediate states instead of  $p$ and
$\bar{p}$; one can read  $f^{p^*}_{p\gamma}$ and
$\bar{f}^{\bar{p}^*}_{\bar p\gamma}$ from equation
(\ref{denomenator}), by setting $\Gamma_{p^*}$ and $\Gamma_{\bar
p^*}$ to zero; moreover
\begin{eqnarray}
\Phi^{\nu}_{p^*}(p_{p}, S_{p}, p_{p^*}, S_{p^*})&=&-e\bar u(p_{p},
S_{p})(\gamma^\nu
-i\frac{\kappa_N}{2M_N}\sigma^{\nu\mu}p_{3\mu})u(p_{p^*},
S_{p^*}),\nonumber\\
\Phi^{C\nu}_{\bar{p}^*}(p_{\bar{p}^*}, S_{\bar{p}^*}, p_{\bar{p}},
S_{\bar{p}} )&=&-e\bar v(p_{\bar{p}^*}, S_{\bar{p}^*})(\gamma^\nu
-i\frac{\kappa_N}{2M_N}\sigma^{\nu\mu}p_{3\mu})v(p_{\bar{p}},
S_{\bar{p}}).\nonumber
\end{eqnarray}
which are obtained from the effective lagrangian of $NN\gamma$.

\section{Conclusion}
To provide a consistent and complete picture of baryon  resonances,
the various possible production and decay channels need to be
explored. With estimated branching ratios for contribution of the
$N^*(1440)$, $N^*(1535)$ and $N^*(1520)$ to the process $J/\psi \to
\gamma p \bar p$, we propose to study radiative decays of excited
nucleon and hyperon states through $J/\psi\to B^*\bar B + \bar B^* B
\to \gamma B\bar B$ processes at BESIII. We provide explicit partial
wave amplitude formulae for these processes with $J^P$ for $B^*$ is
${1\over 2}^{\pm}$, ${3\over 2}^{\pm}$, ${5\over 2}^{\pm}$, ${7\over
2}^{\pm}$. These formulae can be used to perform partial wave
analysis of forthcoming high statistics data from BESIII on these
channels to extract various useful information on the excited
baryons. The BESIII  can produce ground state
($N$,$\Lambda$,$\Sigma$,$\Xi$) and the excited baryon states 
($N^*$,$\Lambda^*$,$\Sigma^*$,$\Xi^*$) via $J/\psi\to B^*\bar B +
\bar B^* B \to \gamma B\bar B$, as well as can do further
investigations into the dynamics of the excited baryons. We hope
that our knowledge about the structure of the excited baryon
resonances  and about the mechanisms of nucleon and hyperon
production will be  clarified by the near future studies at BESIII.

\bigskip
{\bf Acknowledgements:}

We thank Jian-Ping Dai, Yan-Ping Huang, Puze Gao, Kai Ma and Andrey
Sarantsev for useful discussions. This work is Supported by the
National Natural Science Foundation of China (Nos. 10965006,
10875133, 10821063, 10635080, 10905059) and by the Chinese Academy
of Sciences Knowledge Innovation Project (Nos. KJCX2-EW-N01) and by
the Ministry of Science and Technology of China (2009CB825200).

\section*{Appendix: The Relation with the Helicity Amplitude}

In this appendix, we discuss the relation between  amplitudes in the
L-S and helicity formalism for $B^* \to B \gamma$. From
Ref.\cite{copley}, the radiative decay width is related to the
helicity amplitudes $A_{1/2}$ and $A_{3/2}$ as
\begin{eqnarray}\label{helicity}
\Gamma_\gamma=\frac{k^2}{4\pi}\frac{m_B}{m_{B^*}}\frac{8}{2J+1}(|A_{3/2}|^2+|A_{1/2}|^2),
\end{eqnarray}
where $k$ is the three momentum of photon, and $J$ is the total spin
of $B^*$. Let us consider that the photon is moving along the $z$
axis, and the photon is right-handed polarized, in other words, the
spin of the photon is along the $z$-axis. $A_{3/2}$ is the
spin-$3/2$ helicity amplitude of the initial $B^*$ in a state with
$|J, 3/2>$ and final $B$ in a state with $|1/2,1/2>$, and $A_{1/2}$
denotes the spin-$1/2$ helicity amplitude of the $B^*$ with $|J,
1/2>$ and final $B$ with $|1/2,-1/2>$.

In  the L-S Scheme the decay amplitude formulae for  $B^* \to B\gamma$ are
\begin{eqnarray}\label{L-S}
\Gamma_\gamma &=& \frac{k}{2\pi}\frac{m_B}{m_{B^*}}\frac{1}{2J+1}\sum_{s_{zB^*}, s_{zB},s_{z\gamma}}|M|^2\nonumber\\
&=&  \frac{k}{2\pi}\frac{m_B}{m_{B^*}}\frac{1}{2J+1}
( |M_{(3/2,1/2,1)}|^2 +  |M_{(1/2,-1/2,1)}|^2 + |M_{(-1/2,1/2,-1)}|^2  +  |M_{(-3/2,-1/2,-1)}|^2)\nonumber\\
&=&  \frac{k}{\pi}\frac{m_B}{m_{B^*}}\frac{1}{2J+1}
(|M_{(3/2,1/2,1)}|^2 +  |M_{(1/2,-1/2,1)}|^2 ).
\end{eqnarray}
By comparing Eq.(\ref{helicity}) with the Eq.(\ref{L-S}), we can have
the relation between the helicity and L-S  amplitudes  as follows
\begin{eqnarray}
A_{3/2}&=& \frac{1}{2k}|M_{(3/2,1/2,1)}|^2 = \frac{1}{2k}|(g_1 M_{1(3/2,1/2,1)} + g_2 M_{2(3/2,1/2,1)} )|^2 ,\\
 A_{1/2}&=&  \frac{1}{2k}|M_{(1/2,-1/2,1)}|^2 = \frac{1}{2k}|(g_1 M_{1(1/2,-1/2,1)} + g_2M_{2(1/2,-1/2,1)} )|^2.
\end{eqnarray}
As an example, now  we calculate the
$M(s_{B^*},s_{B},s_{\gamma})$ for  $B^*(\frac{3}{2}^+) \to
B\gamma$. From Eq.(\ref{amp}), the two dependent amplitudes can be
written as
\begin{eqnarray}
M_1&=&i \phi^{(1)}_{\mu}\epsilon^{\mu\nu\lambda\sigma}
\tilde{t}^{(1)}_{\lambda}\hat p_{B^*\sigma}\epsilon^*_{\nu}\;,\\
M_2&=&-\frac{2}{3}(\tilde{r}\cdot\tilde{r})\Phi^{(2)\nu\mu}\tilde{t}^{(1)}_{\mu}\epsilon^*_{\nu}+\Phi^{(2)}_{\mu\lambda}\tilde{t}^{(3)\mu\lambda\nu}\epsilon^*_{\nu}\;.
\end{eqnarray}
Before starting our calculation of $M(s_{B^*},s_{B},s_{\gamma})$, we should define
wave functions of particles
\begin{eqnarray}
\bar{u}(p_B,1/2)&=&\sqrt{\frac{E_B+m_B}{2m_B}}(1,0,\frac{k}{E_B+m_B},0),\\
\bar{u}(p_B,-1/2)&=&\sqrt{\frac{E_B+m_B}{2m_B}}(0,1,0,\frac{-k}{E_B+m_B}),\\
u(m_{B^*},1/2)&=&\left(\begin{array}{c}
1 \\ 0 \\ 0 \\ 0 \\\end{array}\right),\hspace{1cm}
u(m_{B^*},-1/2)= \left(\begin{array}{c} 0 \\ 1 \\ 0 \\ 0
\\\end{array}\right),
\end{eqnarray}
and
\begin{eqnarray}
\epsilon^*(p_{\gamma},1,1)&=&(0,-\frac{\sqrt{2}}{2},\frac{\sqrt{2}}{2}i,0),\hspace{0.5cm}
\epsilon(m_{B^*},1,1)=(0,-\frac{\sqrt{2}}{2},-\frac{\sqrt{2}}{2}i,0),\\
\epsilon(m_{B^*},1,0)&=&(0,0,0,1),\hspace{1.7cm}
\epsilon(m_{B^*},1,-1)=(0,\frac{\sqrt{2}}{2},-\frac{\sqrt{2}}{2}i,0),
\end{eqnarray}
where $B^*$ is at rest, in
other words, $p_{B^*}=(m_{B^*},0,0,0)$, $p_{B}=(E_{B},0,0,-k)$ and
$p_{\gamma}=(k,0,0,k)$. By using Eqs.(\ref{epn}-\ref{un}),
the different states of $\phi^{(1)}_{\mu}$ and
$\Phi^{(2)}_{\mu\nu}$ can be obtained
\begin{eqnarray}
\phi^{(1)}_{\mu}(s_{B^*}=1/2,s_{B}=-1/2)&=&\sqrt{\frac{E_B+m_B}{2m_B}}\sqrt{\frac{1}{3}}\epsilon_{\mu} (m_{B^*},1,1),\\
\phi^{(1)}_{\mu}(s_{B^*}=3/2,s_{B}=1/2)&=&\sqrt{\frac{E_B+m_B}{2m_B}}\epsilon_{\mu} (m_{B^*},1,1),\\
\Phi^{(2)}_{\mu\nu}(s_{B^*}=1/2,s_{B}=-1/2)&=&\sqrt{\frac{E_B+m_B}{2m_B}}\sqrt{3}(\epsilon_{\mu} (m_{B^*},1,1)\epsilon_{\nu} (m_{B^*},1,0)\nonumber\\
&&\ \ \ \ +\epsilon_{\mu} (m_{B^*},1,0)\epsilon_{\nu} (m_{B^*},1,1)),\\
\Phi^{(2)}_{\mu\nu}(s_{B^*}=3/2,s_{B}=1/2)&=&-\sqrt{\frac{E_B+m_B}{2m_B}}(\epsilon_{\mu} (m_{B^*},1,1)\epsilon_{\nu} (m_{B^*},1,0)\nonumber\\
&&\ \ \ \ +\epsilon_{\mu} (m_{B^*},1,0)\epsilon_{\nu}
(m_{B^*},1,1)).
\end{eqnarray}
At last, we get the following amplitudes $M(s_{B^*},s_{B},s_{\gamma})$
\begin{eqnarray}
M_1(1/2,-1/2,1)&=&\frac{2k}{\sqrt{3}}\sqrt{\frac{E_B+m_B}{2m_B}},\\
M_1(3/2,1/2,1)&=&2k\sqrt{\frac{E_B+m_B}{2m_B}},\\
M_2(1/2,-1/2,1)&=&\frac{-24k^3}{\sqrt{3}}\sqrt{\frac{E_B+m_B}{2m_B}},\\
M_2(3/2,1/2,1)&=&8k^3\sqrt{\frac{E_B+m_B}{2m_B}},
\end{eqnarray}
where we have used the following relations
\begin{eqnarray}
i\epsilon_{\mu abc}&=& \gamma_5(\gamma_\mu \gamma_a \gamma_b \gamma_c -\gamma_\mu \gamma_a g_{bc} +\gamma_\mu \gamma_b g_{ac} -\gamma_\mu\gamma_c g_{ab}\nonumber\\&& -\gamma_a\gamma_b g_{\mu c} +\gamma_a\gamma_c g_{\mu b} -\gamma_b\gamma_c g_{\mu a} + g_{\mu a} g_{bc} - g_{\mu b }g_{ac} + g_{\mu c}g_{ab}),
\end{eqnarray}
\begin{equation}
\gamma^5=\left(\begin{array}{cc}
0 & \sigma_0 \\ \sigma_0 & 0 \end{array}\right),\hspace{0.5cm}
\gamma^0 = \left(\begin{array}{cc}
I & 0 \\ 0 & -I \end{array}\right),\hspace{0.5cm}
\gamma^i = \left(\begin{array}{cc} 0 & \sigma_i \\
-\sigma_i & 0 \end{array}\right)\;,\hspace{0.5cm} i=1,2,3.
\end{equation}
Now we list the relation between the square of the helicity amplitudes and square of the
coupling constants from our amplitudes for $B^*({1\over 2}^{\pm}$,
${3\over 2}^{\pm}$, ${5\over 2}^{\pm}$, ${7\over 2}^{\pm}) \to
B\gamma$ as follows
\begin{eqnarray}
&&M=g^a_{\gamma}M_1+g^b_{\gamma}M_1,\\
B^*({1\over 2}^+):&&\nonumber\\
 M1&=&i\Phi^{(1)}_\mu\epsilon^{\mu\nu\lambda\sigma}\epsilon^{*}_{\nu}\tilde{t}^{(1)}_{\lambda}\hat p_{B^*\sigma},\\
 |A_{1/2}|^2 &=&  \frac{E_B + m_B}{2m_B}\; 4k|g^a_{\gamma}|^2.\\
B^*({1\over 2}^-):&&\nonumber\\
 M1&=&-\frac{2}{3}(\tilde{r}\cdot\tilde{r})\Phi^{(1)}_{\mu}\epsilon^{*\mu}+\Phi^{(1)}_{\mu}\epsilon^{*}_{\nu}\tilde{t}^{(2)\mu\nu},\\
| A_{1/2}|^2&=&   \frac{E_B+m_B }{m_B}\; 16k|g^a_{\gamma}|^2 .\\
B^*({3\over 2}^+):&&\nonumber\\
 M_1&=&i\phi^{(1)}_{\mu}\epsilon^{\mu\nu\lambda\sigma}\tilde{t}^{(1)}_{\lambda}\hat p_{B^*\sigma}\epsilon^*_{\nu}\;,\\
 M_2&=&-\frac{3}{5}(\tilde{r}\cdot\tilde{r})\Phi^{(2)\mu\nu}\epsilon^*_{\mu}\tilde{t}^{(1)}_{\nu}+\Phi^{(2)}_{\mu\lambda}\tilde{t}^{(3)\mu\lambda\nu}\epsilon^*_{\nu},\\
 |A_{1/2}|^2 &=& \frac{E_B+m_B}{2m_B}\;\frac{2}{3}k|g^a_{\gamma}-12g^b_{\gamma}k^2|^2,\\
 |A_{3/2}|^2&=&\frac{E_B+m_B}{2m_B}\;2k|g^a_{\gamma}+ 4 g^b_{\gamma}k^2|^2.\\
B^*({3\over 2}^-):&&\nonumber\\
 M_1&=&-\frac{2}{3}(\tilde{r}\cdot\tilde{r})\phi^{(1)\mu}\epsilon^*_{\mu}+\phi^{(1)}_{\mu}\tilde{t}^{(2)\mu\nu}\epsilon^*_{\nu},\\
 M_2&=&i\Phi^{(2)}_{\mu\alpha}\epsilon^{\mu\nu\lambda\sigma}\epsilon^*_{\nu}\tilde{t}^{(1)\;\alpha}_{\lambda}\hat p_{B^*\sigma}\;,\\
 |A_{1/2}|^2 &=& \frac{E_B+m_B}{2m_B}\;\frac{8}{3}k^3|-g^a_{\gamma}+3g^b_{\gamma}|^2,\\
 |A_{3/2}|^2&=&\frac{E_B+m_B}{2m_B}\;8k^3|-g^a_{\gamma}- g^b_{\gamma}|^2.\\
B^*({5\over 2}^+):&&\nonumber\\
 M_1&=&-\frac{3}{5}(\tilde{r}\cdot\tilde{r})\phi^{(2)\mu\nu}\epsilon^*_{\mu}\tilde{t}_{(1)\nu}+\phi^{(2)}_{\mu\nu}\tilde{t}^{(3)\mu\nu\lambda}\epsilon^*_{\lambda},\\
 M_2&=&i\Phi^{(3)}_{\mu\alpha\beta}\epsilon^{\mu\nu\lambda\sigma}\epsilon^*_{\nu}\tilde{t}^{(3)\;\alpha\beta}_{\lambda}\hat p_{B^*\sigma}\;,\\
|A_{1/2}|^2 &=& \frac{E_B+m_B}{2m_B}\;\frac{32}{5}k^5|-g^a_{\gamma} + 4g^b_{\gamma}|^2,\\
 |A_{3/2}|^2&=&\frac{E_B+m_B}{2m_B}\;\frac{64}{5}k^5|-g^a_{\gamma} - 2g^b_{\gamma}|^2.\\
B^*({5\over 2}^-):&&\nonumber\\
 M_1&=&i\phi^{(2)}_{\mu\alpha}\epsilon^{\mu\nu\lambda\sigma}\epsilon^*_{\nu}\tilde{t}^{(2)\ \alpha}_{\lambda}\hat p_{B^*\sigma}\;,\\
 M_2&=&-\frac{4}{7}(\tilde{r}\cdot\tilde{r})\Phi^{(3)\mu\nu\lambda}\epsilon^{*}_{\mu}\tilde{t}^{(2)}_{\nu\lambda}+\Phi^{(3)}_{\mu\nu\lambda}\tilde{t}^{(4)\mu\nu\lambda\sigma}\epsilon^*_{\sigma},\\
 |A_{1/2}|^2 &=& \frac{E_B+m_B}{2m_B}\;\frac{8}{5}k^3|g^a_{\gamma}-16g^b_{\gamma}k^2|^2,\\
 |A_{3/2}|^2&=&\frac{E_B+m_B}{2m_B}\;\frac{16}{5}k^3|-g^a_{\gamma}+ 4g^b_{\gamma}k^2|^2.\\
B^*({7\over 2}^+):&&\nonumber\\
 M_1&=&i\phi^{(3)}_{\mu\alpha\beta}\epsilon^{\mu\nu\lambda\sigma}\epsilon^*_{\nu}\tilde{t}^{(3)\ \alpha\beta}_{\lambda}\hat p_{B^*\sigma}\;,\\
 M_2&=&-\frac{5}{9}(\tilde{r}\cdot\tilde{r})\Phi^{(4)\mu\nu\lambda\sigma}\epsilon^{*}_{\mu}\tilde{t}^{(3)}_{\nu\lambda\sigma}+\Phi^{(4)}_{\mu\nu\lambda\sigma}\tilde{t}^{(5)\mu\nu\lambda\sigma\delta}\epsilon^*_{\delta},\\
  |A_{1/2}|^2 &=& \frac{E_B+m_B}{2m_B}\;\frac{128}{35}k^5|g^a_{\gamma}-20g^b_{\gamma}k^2|^2,\\
 |A_{3/2}|^2&=&\frac{E_B+m_B}{2m_B}\;\frac{128}{21}k^5|g^a_{\gamma}+ 12g^b_{\gamma}k^2|^2.\\
B^*({7\over 2}^-):&&\nonumber\\
 M_1&=&-\frac{4}{7}(\tilde{r}\cdot\tilde{r})\phi^{(3)\mu\nu\lambda}\epsilon^*_{\mu}\tilde{t}_{(1)\nu\lambda}+\phi^{(3)}_{\mu\nu\lambda}\tilde{t}^{(4)\mu\nu\lambda\sigma}\epsilon^*_{\sigma},\\
 M_2&=&i\Phi^{(4)}_{\mu\alpha\beta\gamma}\epsilon^{\mu\nu\lambda\sigma}\epsilon^*_{\nu}\tilde{t}^{(4)\;\alpha\beta\gamma}_{\lambda}\hat p_{B^*\sigma}\;,\\
  |A_{1/2}|^2 &=& \frac{E_B+m_B}{2m_B}\;\frac{512}{35}k^7|-g^a_{\gamma}+5g^b_{\gamma}|^2,\\
 |A_{3/2}|^2&=&\frac{E_B+m_B}{2m_B}\;\frac{512}{21}k^7|-g^a_{\gamma}-3g^b_{\gamma}|^2.
\end{eqnarray}

\end{document}